\documentclass[prb,twocolumn,showpacs,aps,superscriptaddress,floatfix]{revtex4-1}
\usepackage{amsmath}
\usepackage{amssymb}
\usepackage{amsfonts}
\usepackage{bm}
\usepackage[dvips]{graphicx}
\usepackage{color}
\usepackage{ulem}
\usepackage[unicode, pdftex]{hyperref}
\usepackage{verbatim}
\DeclareGraphicsExtensions{.eps,.pdf}
\graphicspath{{pictures/}}
\usepackage{epstopdf}
\begin{document}
	\title{Quasiparticle interference in doped topological insulators with nematic superconductivity}
    \author{D. A. Khokhlov}
    \affiliation{Dukhov Research Institute of Automatics, Moscow, 127055 Russia}
    \affiliation{Moscow Institute of Physics and Technology, Dolgoprudny,
    Moscow Region, 141700 Russia}
    \affiliation{National Research University Higher School of Economics, 101000 Moscow, Russia}
   
    \author{R. S. Akzyanov}
    \affiliation{Dukhov Research Institute of Automatics, Moscow, 127055 Russia}
       \affiliation{Moscow Institute of Physics and Technology, Dolgoprudny,
    Moscow Region, 141700 Russia}
    \affiliation{Institute for Theoretical and Applied Electrodynamics, Russian
    Academy of Sciences, Moscow, 125412 Russia}

	\begin{abstract}
        We theoretically investigate quasiparticle interference in superconducting topological insulators with the nematic order parameter. This order parameter spontaneously breaks the rotational symmetry of the crystal. Such rotational symmetry breaking is visible in the quasiparticle interference picture both in coordinate and momentum spaces. For a small bias voltages quasiparticle interference incommensurate with the crystal symmetry and shows nematic behavior. If the bias voltage is comparable with the value of the order parameter interference picture is similar to the interference picture of the normal state. Interference patterns are sensitive to the orientation of the nematicity. We compare our results with the existing experimental data. 
	\end{abstract}
	\date{\today}
	
	\maketitle
	\section{Introduction}

In recent years, bulk superconductivity in doped topological insulators, such as A$_x$Bi$_2$Se$_3$ (A stands for Nb, Cu or Sr), attracts significant attention~\cite{Sasaki2011,Chen2018, Charpentier2017,Du2017,Yonezawa2016,Kasahara2012,Asaba2017,Li2018,Venderbos2018,Brydon2014,Hecker2018,Chiba2017,Hor2010,Kirzhner2012,Kriener2011,Kuntsevich2018,Kuntsevich2019,Kozii2015}. Measurement of the Knight shift verify the spin-triplet origin of the superconductivity in these materials~\cite{Matano2016}. Contact measurements reveal that this superconductivity show non-BCS behaviour~\cite{Kirzhner2012,Tao2018,Sirohi2018,Sasaki2011}. The second critical field has the two-fold in-plane rotational symmetry that is inconsistent with the three-fold rotational crystal symmetry of Bi$_2$Se$_3$\cite{Pan2016,Kuntsevich2019}. Measurements of the magnetic torque in Nb-doped Bi$_2$Se$_3$ show two-fold in-plane symmetry as well~\cite{Asaba2017}. This rotational symmetry breaking indicates the emergence of the nematic order with the triplet pairing in the system\cite{Fu2010,Fu2014,Venderbos2016}.

Theoretical calculations show that nematic superconducting order with E$_{u}$ representation that spontaneously breaks inversion symmetry is possible in topological insulators~\cite{Fu2010}. This order parameter is a two-component vector~\cite{Fu2014,Kawai2020}. The orientation of the vector is associated with the direction of the nematicity that affects the physical properties of the system such as anisotropy of the second critical field\cite{Venderbos2016}. Experiments show that in different compounds orientation of the nematicity can be parallel~\cite{Tao2018,Matano2016,Andersen2018} or perpendicular~\cite{Chen2018,Yonezawa2016} to the main crystal direction [001] that refers to $\Delta_{4x}$ and $\Delta_{4y}$ pairings correspondingly~\cite{Yonezawa2018}. Moreover, in multiblock samples, different orientations of the nematicity in different domains are realized~\cite{Kostylev2020}. 

In Bi$_2$Se$_3$ presence of the third order in momentum anisotropic terms leads to Fermi surface with hexagonal deformation~\cite{Kuroda2010}. These terms are referred to as hexagonal warping~\cite{Fu2009}. Such warping has a significant effect on the properties of the topological insulators~\cite{Akzyanov2018,Akzyanov2019} and particularly on the nematic superconducting state. Namely, it opens a full gap in the spectrum if nematicity is not aligned along one of the six main crystal axis~\cite{Fu2014}. 
Generally speaking, in the presence of hexagonal warping different orientations of the nematicity becomes non-equivalent.

One of the direct ways to observe electronic structure in the experiment is the quasiparticle interference~\cite{Avraham2018,Hasan2010} (QPI) using scanning tunneling microscopy (STM). A probe of the STM measures the spatial variation of the local density of states due to the interference of the electrons on the impurities. Fourier transform of the local density of states contains information about the scattering vectors that provide us insights about the electronic structure of the material. In superconductors, QPI has become a powerful tool for elucidating the nature of the quasiparticle states in novel superconductors~\cite{Akbari2010,Hirschfeld2015,Lee2010,Farrell2015,Boker2019,Chaocheng2017,Iwaya2017,Gu2018,Wang2017,Gu2018}. 

In the recent experiment QPI in the Bi$_2$Te$_3$ film placed on the iron-based superconductor FeTe$_{0.55}$Se$_{0.45}$ has been measured~\cite{Chen2018}. Superconductivity is induced in the thin film of Bi$_2$Te$_3$ via the proximity effect. STM measurements reveal that at bias voltages exceeding the gap value, QPI consists of the single hexagon at large momentum. This interference pattern is similar to the QPI of the normal state~\cite{Zhou2009} of Bi$_2$Te$_3$. For smaller values of the voltage, only two sides of the hexagon remain in the opposite directions. This twofold symmetry breaks the rotational symmetry of the normal state and arises due to the nematic superconductivity of the system. 

In our work, we theoretically investigate QPI of Bi$_2$Se$_3$
with the nematic superconductivity with $E_u$ symmetry of the order parameter. We use low energy Hamiltonian of the bulk states 
of Bi$_2$Se$_3$ from Ref.~\onlinecite{Liu2010}. We consider two different orientations of the nematic order parameter that corresponds to the $\Delta_{4x}$ and $\Delta_{4y}$ pairings. We calculate QPI in both real and reciprocal spaces due to the scattering of the impurity using the T-matrix formalism. We found that QPI shows nematic behavior if the bias voltage is smaller than the value of the order parameter. This nematic behaviour is visible for both short-wave and long-wave scattering vectors. The difference between the interference patterns for the different orientations of the nematicity is visible for the shortwave scattering. Difference between different orientations of the nematicity is prominent in the coordinate space. We compare our results with the experiment of Bi$_2$Te$_3$/FeTe$_{0.55}$Se$_{0.45}$ from Ref.~\onlinecite{Chen2018}. In order to match our results with the experimental data we consider low energy model of the normal state Hamiltonian with only single warping parameter. We found that both short-wave and long-wave QPI from our calculations are similar to the experimental data. Also, we checked that surface Andreev bound states do not contribute to QPI.


The paper is organized as follows: in Sec.~\ref{sec_model} we describe topological insulator in normal state and nematic superconducting state. In Sec.~\ref{sec_method} method of QPI calculation is presented. Results of the QPI calculations are discussed in Sec.~\ref{sec_qpi}. We give a comparison with the experiment in Sec.~\ref{sec_exp}. Discussion of the obtained results and summarized conclusions are presented in Sec.~\ref{sec_discussion}.   
	\section{Model}\label{sec_model}
	\subsection{Normal state}
Low energy Hamiltonian of the bulk states in Bi$_2$Se$_3$ is described in several papers~\cite{Zhang2009, Liu2010}. Hamiltonian of the bulk states of the topological insulator is written as 
    \begin{eqnarray}
        \label{Eq::disorder_ham_reduce}
        \nonumber
        H_0(\mathbf{k})=
        -\mu +
        m\sigma_z+
        v(s_x\sigma_xk_y-s_y\sigma_xk_x)+
        v_zs_x\sigma_y+\\
        \lambda_1(k_x^3-3k_xk_y^2)s_z\sigma_x +
        \lambda_2(k_y^3-k_yk_x^2)\sigma_x.\quad
    \end{eqnarray}  
    Here Pauli matrices $s_{x,y,z}$ acts in spin space and $\sigma_{x,y,z}$ acts in orbital space. Momentum in $(x,y,z)$ directions are doentoed as $(k_x,k_y,k_z). $Fermi velocities in $xy$ plane and $z$ directions are $v$ and $v_z$, $\mu$ is the chemical potential, $\lambda_1$ and $\lambda_2$ defines two different hexagonal warpings. Parameter $m$ describes the single electron gap between electron and valence bands. We omit quadratic corrections to the spectra since they do not bring any new sufficient physics. In further consideration we neglect dispersion along $z$ direction since STM is the surface technique and only the states propagating in $(x,y)$ plane contribute to the QPI. 
    \subsection{Superconducting state}
    We work in a Nambu basis
    \begin{eqnarray}
        \label{Eq::wave_function}
        \Psi_{\mathbf{k}}=(\phi_{\mathbf{k}},-is_y\phi^{\dagger}_{-\mathbf{k}})^t,
    \end{eqnarray}
    where $\phi_{\mathbf{k}}=(\phi_{\uparrow,1,\mathbf{k}},\phi_{\downarrow,1,\mathbf{k}},\phi_{\uparrow,2,\mathbf{k}},\phi_{\downarrow,2,\mathbf{k}})^t$. Here $\uparrow(\downarrow)$ means up (down) projection of the spin on z-axis, $1,2$ corresponds to the different orbitals and the superscript $t$ means transposition. In this basis topological insulator with the nematic superconductivity is described by the $8\times8$ BdG Hamiltonian~\cite{Hao2017}:
    \begin{eqnarray}
    \label{Eq::sc_ham}
        H_{\text{BdG}}(\mathbf{k})= H_0(\mathbf{k})\tau_z+\hat{\Delta}\tau_x,
    \end{eqnarray}
    where Pauli matrices $\tau_{x,y,z}$ act in electron-hole space, $\hat{\Delta}=\Delta\sigma_y{\mathbf s}\cdot {\mathbf n}$ is the superconducting order parameter. Here ${\mathbf s}=(s_x,s_y)$ and $\mathbf{n}=(\cos \alpha;\sin{\alpha})$ is a real unit vector that shows direction of the nematicity of the nematic order parameter within $E_u$ symmetry~\cite{Fu2010}. This order parameter is a vector $\Delta=(\Delta_{4x},\Delta_{4y})$. Orientation of this vector is determined by combination of the strain and warping terms~\cite{Fu2014}. We consider two orientations of the nematicity $\alpha=0$ and $\alpha=\pi/2$ that corresponds to the pairings $\Delta_{4x}$ and $\Delta_{4y}$ respectively.
    
Without warpings $\lambda_1=\lambda_2=0$, the spectrum of the Hamiltonian given by Eq.~\ref{Eq::sc_ham} has two nodes. Warping term $\lambda_1$ opens the full gap for all nematicity orientations besides $\Delta_{4x}$~\cite{Fu2014}. If we include another warping $\lambda_2$ then full gap opens for all orientations of the nematicity for $k_z=0$. The full gap for the states propagating in $k_z=0$ plane has been measured in tunneling measurements~\cite{Tao2018,Matano2016,Andersen2018}. Note, that for 3D Hamiltonian nodes are present for $\Delta_{4x}$ for some $k_z\neq 0$~\cite{Fu2014}.

	\section{Method}
	\label{sec_method}
Quasiparticles interfere on the impurities or defects in the sample. Such interference results in local oscillations of the quasiparticle density in the real space $\rho(\mathbf{r},\omega)$ that can be measured using STM tip at different biases $\omega$. Fourier transform of the $\rho(\mathbf{r},\omega)$ gives quasiparticle interference picture in momentum space $\rho(\mathbf{q},\omega)$.  
We consider point charged impurity with the potential
	\begin{eqnarray}
        \label{Eq::perturbation}
        V(\mathbf{r})=V\delta(\mathbf{r}),
    \end{eqnarray}
              \begin{figure*}[htbp!]
        \center{\includegraphics[width=0.8\linewidth]{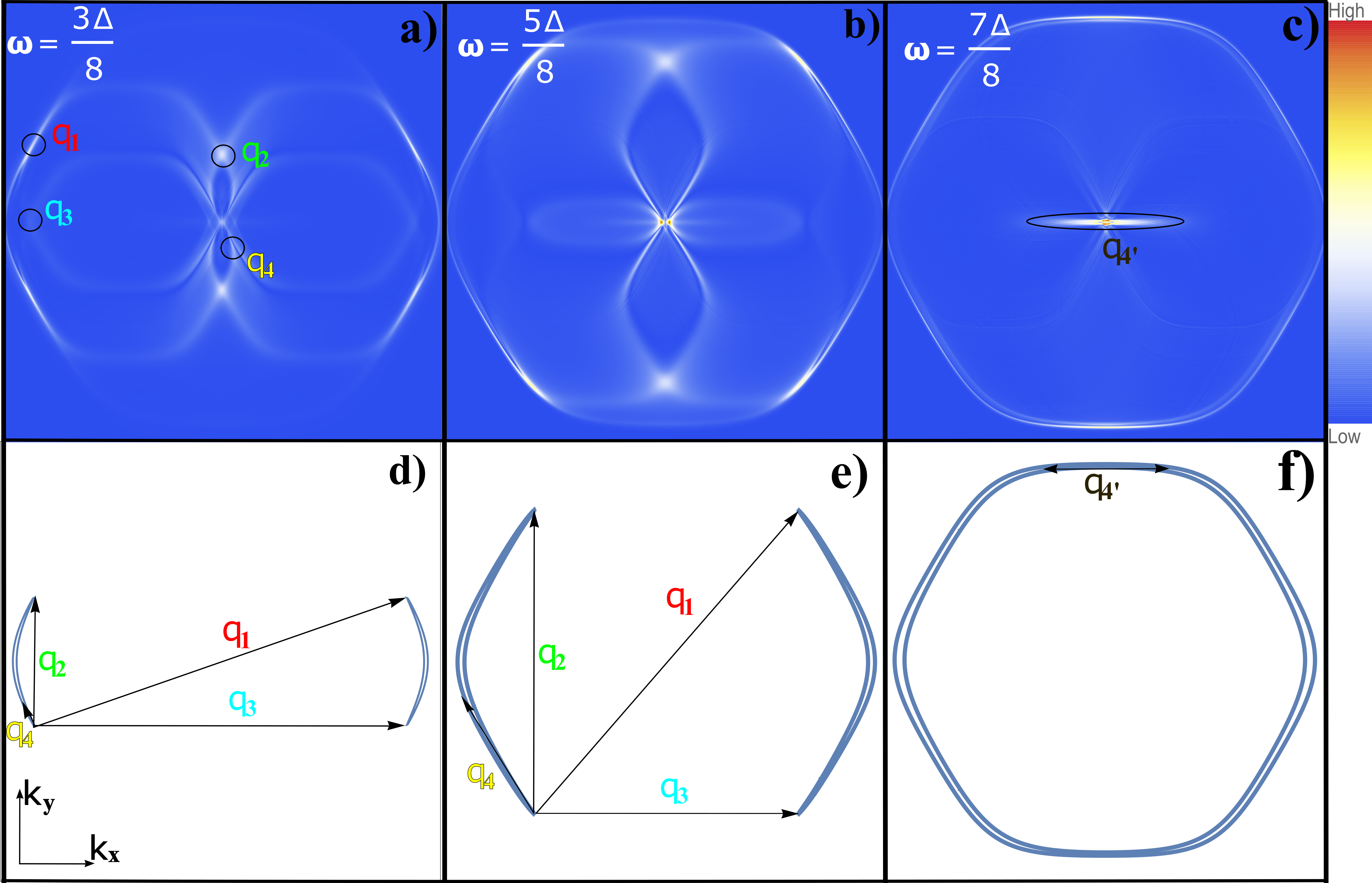}}
        \caption{QPI in momentum space for nematicity orientation $\Delta_{4y}$ is shown on panels~a-c for different values of the bias voltages $\omega$. Panels~d-f: constant energy contours at the same voltages. Different scattering channels are marked by circles in panels~a,~c, and by arrows in panels~d-f. Note, that scattering vectors in panels d-f are twice as long than these vectors in panels a-c.} 
        \label{Fig::QPI&FS_alpha_pi}
    \end{figure*}
    \begin{figure*}[htbp!]
            \center{\includegraphics[width=0.8\linewidth]{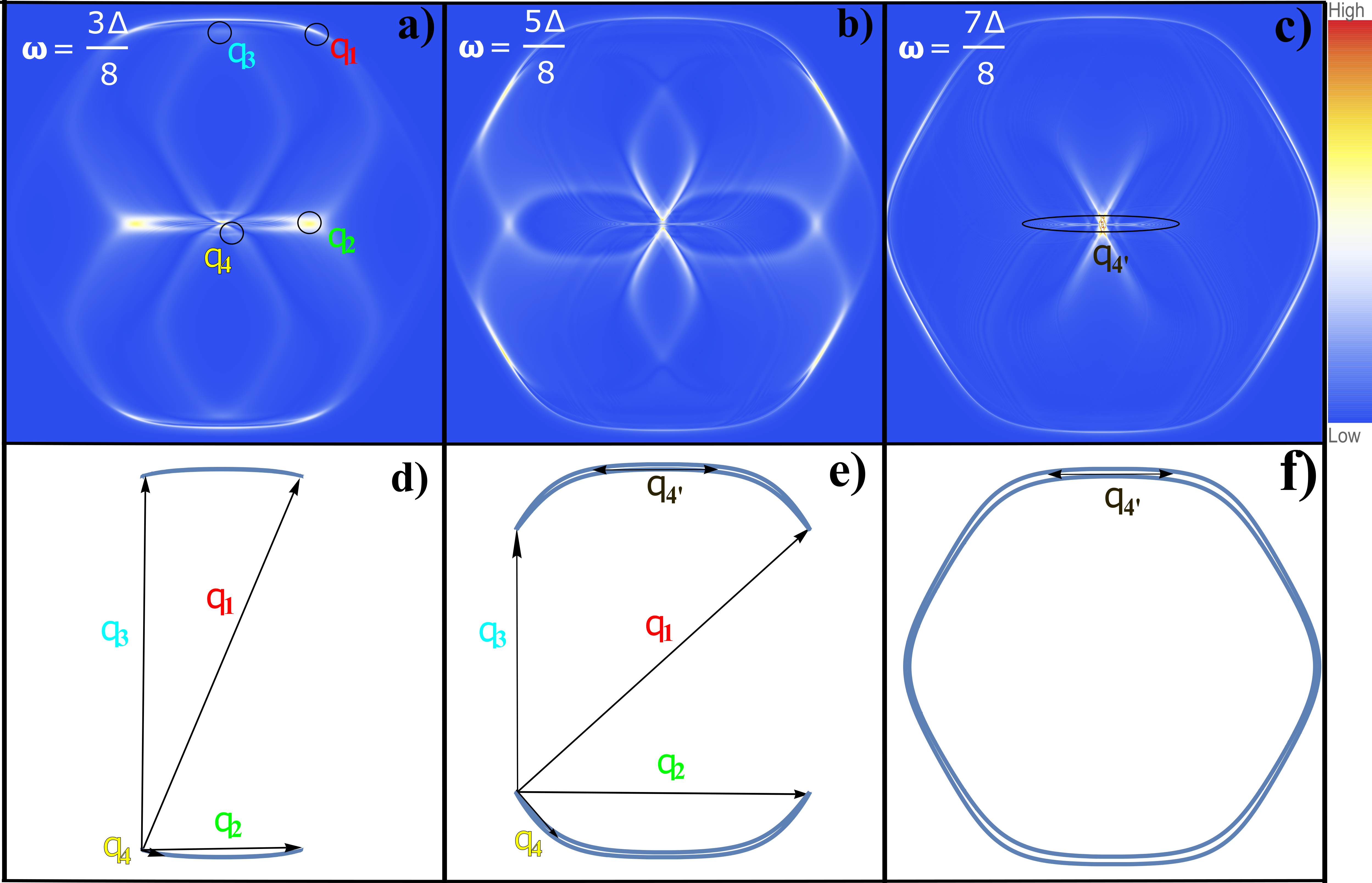}}
        \caption{QPI in momentum space for nematicity orientation $\Delta_{4x}$ is shown on panels~a-c for different values of the bias voltages $\omega$. Panels~d-f: constant energy contours at the same voltages. Different scattering channels are marked by circles in panels~a,~c, and by arrows in panels~d-f. Note, that scattering vectors in panels d-f are twice as long than these vectors in panels a-c.} 
        \label{Fig::QPI&FS_alpha_0}
    \end{figure*}
	where $V=V_0\tau_z$. Here $V_0$ is the strength of the impurity potential and $V(\mathbf{r})$ acts in the space of 8-component Nambu spinors. Note, that charged impurity acts differently on electrons and on holes thus we get nontrivial matrix structure $\tau_z$ of the scatter. For numerical calculations we take $V_0=20\;\text{eV}\cdot\text{\AA}^2$ that corresponds to the typical value of the charged impurity~\cite{Beidenkopf2011}. 
    
    Local density of states is defined by the following equation
    \begin{eqnarray}
        \label{Eq::rho_real}
        \rho(\mathbf{r},\omega)=\frac{1}{2}\textrm{Tr}[(1+\tau_z)\,G(\mathbf{r},\mathbf{r},\omega)],
    \end{eqnarray}
    where factor $(1+\tau_z)/2$ counts contribution from the electrons to the density of states only and disregards contribution of the holes. After Fourier transform of $\rho(\mathbf{r},\omega)$ we get QPI in momentum space
    \begin{eqnarray}
        \label{Eq::rho_momentum}
        \rho(\mathbf{q},\omega)\!=\!\frac{i}{2\pi}\!\sum_{\mathbf{k}}\!\textrm{Tr}[G(\mathbf{k},\mathbf{k}\!-\!\mathbf{q},\omega)\!-\!G^*(\mathbf{k},\mathbf{k}+\mathbf{q},\omega)],\,\,
    \end{eqnarray}
    where notation $^*$ means complex conjugation. 
   We start from the Green's function of the non-perturbed system 
    \begin{eqnarray} 
    G_0(\mathbf{k},\omega)=\left(i\delta+\omega-H_{\text{BdG}}(\mathbf{k})\right)^{-1}.
    \label{Eq::Green_non_perturebed}
    \end{eqnarray}
 We calculate Green's function $G(\mathbf{k},\mathbf{k}',\omega)$ through the T-matrix approach  
 \begin{eqnarray} 
    G(\mathbf{k},\mathbf{k}'\omega)\!\!=\!\!G_0(\mathbf{k},\omega)\delta_{\mathbf{k},\mathbf{k}'}\!\!+\!\!G_0(\mathbf{k},\omega)T(\omega)G_0(\mathbf{k}',\omega),
    \end{eqnarray}
    where $T(\omega)$ is the T-matrix arises due to impurity scattering. For point scatter T-matrix is written exactly as
    \begin{eqnarray}
    T(\omega)=(1-V\sum_{\mathbf{p}}G_0(\mathbf{p},\omega))^{-1}V.
    \end{eqnarray}
Integral that is given by Eq.~\ref{Eq::rho_momentum} is a convolution. 
We can significantly speed up calculation using Fast Fourier transform in comparison with the direct calculation on a grid. We find correction to the Green's function due to scattering in the real space via convolution theorem
    \begin{eqnarray}
    \delta G(\mathbf{r},\mathbf{r}',\omega)=\text{FFT}[G_0(\mathbf{k},\omega)]T(\omega)\text{FFT}[G_0(\mathbf{k},\omega)],
    \label{Eq::exact_GF_add_RS}
    \end{eqnarray}
    where $\text{FFT}[..]$ means Fast Fourier Transform. We are interesting only in correction to the electron density $\delta\rho(\mathbf{r},\omega)=1/2\text{Tr}[\delta G(\mathbf{r},\mathbf{r}',\omega)(1+\tau_z)]$ that occurs due to scattering. Then we find correction to the density in the momentum space as 
    \begin{eqnarray}    
        \delta\rho(\mathbf{q},\omega)\!=\!\frac{1}{2i}\!\left(\text{FFT}^{-1}[\rho(\mathbf{r},\omega)]\!-\!\text{FFT}^{-1*}[\rho(\mathbf{-r},\omega)]\right).\quad
    \end{eqnarray}
In context of T-matrix formalism for the quasiparticle scattering this method was described in details in the Appendix A of the Ref.~\onlinecite{Kohsaka2017}.
    \section{Quasiparticle interference}
    \label{sec_qpi}
\begin{figure*}
        \center{\includegraphics[width=0.8\linewidth]{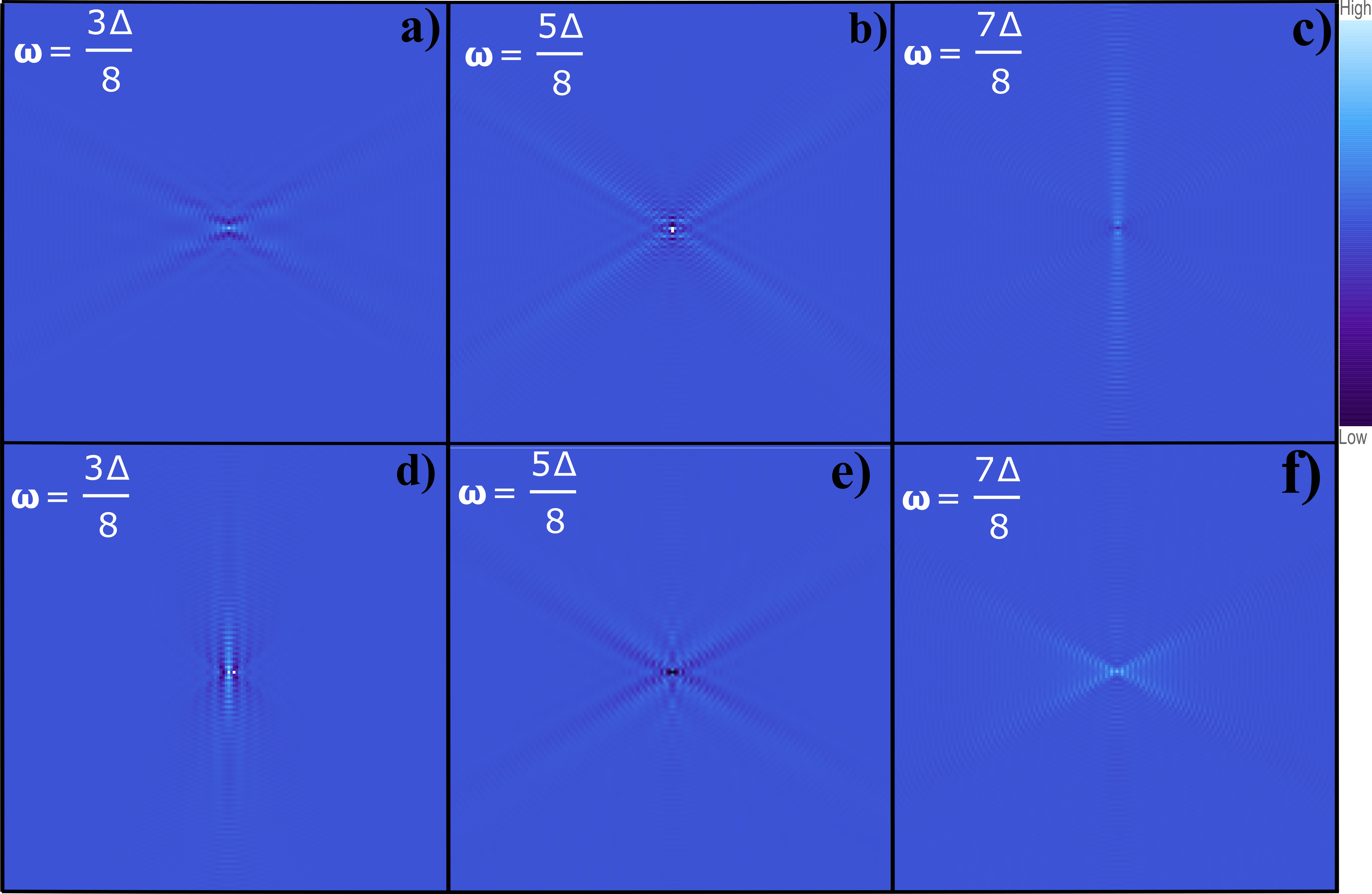}}
        \caption{QPI in the real space. Panels~a-d shows QPI for nematicity orientation $\Delta_{4y}$ at different bias voltages. Panels~e-h shows QPI for nematicity orientation $\Delta_{4x}$ at the same bias voltages. } 
	    \label{Fig::QPI_RS}
    \end{figure*}

    We calculate QPI in real $\delta\rho(\mathbf{r},\omega)$ and momentum spaces $|\delta\rho(\mathbf{q},\omega)|$ at several bias voltages $\omega$. QPI patterns for the nematicity orientations $\Delta_{4x}$ and $\Delta_{4y}$ are presented in Figs.~\ref{Fig::QPI&FS_alpha_0}a-d and~\ref{Fig::QPI&FS_alpha_pi}a-d correspondingly. QPI patterns are accompanied by constant energy contours that are placed under the corresponding QPI, see Figs.~\ref{Fig::QPI&FS_alpha_pi}e-h and~\ref{Fig::QPI&FS_alpha_0}e-h. In addition, we calculate QPI patterns in the real space that are presented in Fig.~\ref{Fig::QPI_RS}. We take the following dimensionless parameters: chemical potential~\cite{Neupane2016} $\mu/m=1.7$, warping constants~\cite{Liu2010} $\lambda_1 m^2/v^3=0.14$,  $\lambda_2 m^2/v^3=0.28$, order parameter $\Delta/m=3\cdot10^{-2}$ and quasiparticle broadening $\delta/m=10^{-3}$. The order parameter and broadening satisfy the following condition $\delta\ll\Delta\ll\mu-m$, where the first inequality corresponds to the clean case and the second one to the weak coupling regime. Values the gap for these parameters are $\overline{\Delta}\approx0.2\Delta$ for nematicity direction $\Delta_{4y}$ and $\overline{\Delta}\approx0.3\Delta$ for nematicity direction $\Delta_{4x}$. 

     Constant energy plots $E_{i{\bf k}}=\omega$ where $E_{i{\bf k}}$ is the energy spectrum of the Hamiltonian consist of two disconnected contours. Symmetry of this contours is governed by time-reversal symmetry. With the increase of the bias $\omega$ contours form a hexagonal structure. At $\omega\gtrsim7/8\Delta$ isoenergy contours consist of two hexagons.  

    While the bias voltage is lower than the value of the gap $\omega<\overline{\Delta}$ there is no QPI since there are no quasiparticles to scatter inside the gap. If the bias voltage is large, $\omega\gtrsim7/8\Delta$, constant energy contours are almost equal for different orientations of the nematicity and consist of two hexagons. 
    
    Nematic properties of QPI are visible if the bias voltage is larger than the value of the gap but smaller than the value of the order parameter $\overline{\Delta}<\omega<7/8\Delta$. In this range of bias energy contour consists of two disconnected contours. We find that there are four main scattering channels with wave vectors $\mathbf{q}_{1,2,3,4}$. These scattering vectors are marked by corresponding arrows at the constant energy contours in Figs.~\ref{Fig::QPI&FS_alpha_0}e-g and~\ref{Fig::QPI&FS_alpha_pi}e-g. Wave vectors are also marked by circles in the corresponding figures in Figs.~\ref{Fig::QPI&FS_alpha_0}a and ~\ref{Fig::QPI&FS_alpha_pi}a. Vectors $\mathbf{q}_{1,3}$ corresponds to intercontour scattering while vectors $\mathbf{q}_{2,4}$ corresponds to intracontour scattering.
    
    The density of states concentrates on the tips of the constant energy contours. Thus, major scattering events are associated with those tips.

   We found that the most prominent long wavevector scattering occurs at $\mathbf{q}_1$ vector. This vector connects the tip of the contour with its time-reversal partner of another contour. Scattering at this vector generates four stripes located at the sides of the hexagon. 
 Vector $\mathbf{q}_1$ corresponds to the back-scattering. In general, scattering process between states $\mathbf{k}$ and $\mathbf{k}'$ is possible only if corresponding matrix element is non-zero $\langle \psi_{\mathbf{k}\alpha}| \hat{V}| \psi_{\mathbf{k}'\beta}\rangle \neq 0$. On the surface of normal topological insulator spectrum is not degenerate and $\alpha=\beta$. In this case back-scattering $\mathbf{k}\to -\mathbf{k}$ occurs between time-reversal symmetric points. For such a pair of states we have $\langle \psi_{\mathbf{k}}|\hat{T}|\psi_{\mathbf{k}}\rangle=0$  and back-scattering is forbidden~\cite{Zhou2009,Lee2009}, where $\hat{T}=is_y K$ and $K$ is complex conjugation. Spectrum of the normal state bulk Hamiltonian given by Eq.~\ref{Eq::disorder_ham_reduce} is doubly degenerate, eigenstates $\phi_{\mathbf{k}I}$ and $\phi_{\mathbf{k}II}$ have energy $E_{0\mathbf{k}}$. Now we have two Kramers pairs and scattering processes between different pairs $\phi_{\mathbf{k}II,I}\to\phi_{-\mathbf{k}I,II}$ are not forbidden. Since back-scattering is possible the normal phase, it is possible in the superconducting state.   

	A significant contribution to QPI occurs due to scattering at $\mathbf{q}_2$ vector. This vector connects tips of the same contour. Scattering at this vector generates two bright points. At high bias $\omega \sim 7/8 \Delta$ those dotes transform into the two stripes that together with $\mathbf{q}_1$ reflexes form a six stripes hexagonal structure.
	
	We found that scattering at $\mathbf{q}_3$ vector, which connects nearby tips of different contours, is suppressed. Short wave vector scattering is determined by the scattering vectors $\mathbf{q}_4$ and $\mathbf{q}_{4'}$. The vector $\mathbf{q}_4$ corresponds to the propagation of the states from the tip along the contour. This process generates two oblique crossed lines at $\mathbf{q}=0$. At larger energies, tips merge and additional scattering channel $\mathbf{q}_{4'}$ occurs. This channel corresponds to the short wavelength scattering at $x$ direction along the side of the hexagon. One may find asymmetry between these stripes, which is generated by nematic asymmetry in the spectrum.
 
	Now, we focus on the differences between the QPI in $\Delta_{4x}$ and $\Delta_{4y}$ superconducting phases. The direction of the gap in $k$-space is perpendicular to the orientation of the order parameter. Phase with $\Delta_{4x}$ orientation has the minimal gap along $k_x=0$, while phase $\Delta_{4y}$ has minimal gap along $k_y=0$. Some distinctions in QPI are generated by the difference of the gap orientation i.e. by the difference in the spectrum. Note that the spectrum of the $\Delta_{4x}$ phase can not be turned to the spectrum of the $\Delta_{4y}$ phase by rotation. This rotational symmetry is broken by hexagonal warping terms.
    
    



    The most prominent difference in QPI between $\Delta_{4x}$ and $\Delta_{4y}$ occurs at the shortwave scattering at $\mathbf{q}_{4}$ and $\mathbf{q}_{4'}$ vectors. As we can see from Figs.~\ref{Fig::QPI&FS_alpha_pi},\ref{Fig::QPI&FS_alpha_0}f, scattering at $\mathbf{q}_{4'}$ vector is possible only at high bias $\omega \sim \Delta$ and only for the $\Delta_{4y}$ orientation of the nematicity. Withal, scattering at the vector $\mathbf{q}_{4}$ at the high bias $\omega \sim \Delta$ is visible only for the $\Delta_{4x}$ orientation. The next difference is the position of the peaks at $\mathbf{q}_{2}$ regarding the position of the $\mathbf{q}_{4}$ stripes at lower biases $\omega \gtrsim 5/8 \Delta$. Scattering at $\mathbf{q}_{4}$ produces two crossed stripes. For $\Delta_{4x}$ orientation scattering at $\mathbf{q}_{2}$ is opposite the obtuse angle of the cross $\mathbf{q}_{4}$ while for $\Delta_{4y}$ orientation scattering at $\mathbf{q}_{2}$ is opposite the acute angle of the cross at $\mathbf{q}_{4}$.

    We plot QPI in the real space for two orientations of the nematicity, see Fig.~\ref{Fig::QPI_RS}a-d for orientation $\Delta_{4y}$ and Fig.~\ref{Fig::QPI_RS}e-h for orientation $\Delta_{4x}$. Friedel oscillations spread along two directions: a vertical stripe that is generated by $\mathbf{q}_2$ scattering and two symmetrical inclined stripes that are generated by $\mathbf{q}_2$ scattering. Note, the angle between these two stripes is not constant.  
    
    For orientation $\Delta_{4y}$ at the bias voltages $3\Delta/8$ and $5\Delta/8$, we find the two inclined stripes, which correspond to intercontour scattering in $\mathbf{q}_1$ channel. They become weak at voltage $\omega=7\Delta/8$ while the vertical stripe, associated with scattering vector $\mathbf{q}_2$, arises. 
    
    For orientation $\Delta_{4x}$ at the voltage $3\Delta/8$, we observe only the vertical stripe, which corresponds to intracontour scattering at the vector $\mathbf{q}_2$. At higher voltages, the inclined stripes associated with vector $\mathbf{q}_1$ arise, while the vertical one disappear. 

\section{Comparison with the experiment}\label{sec_exp}
\begin{figure}[h!]
    \center{\includegraphics[width=1\linewidth]{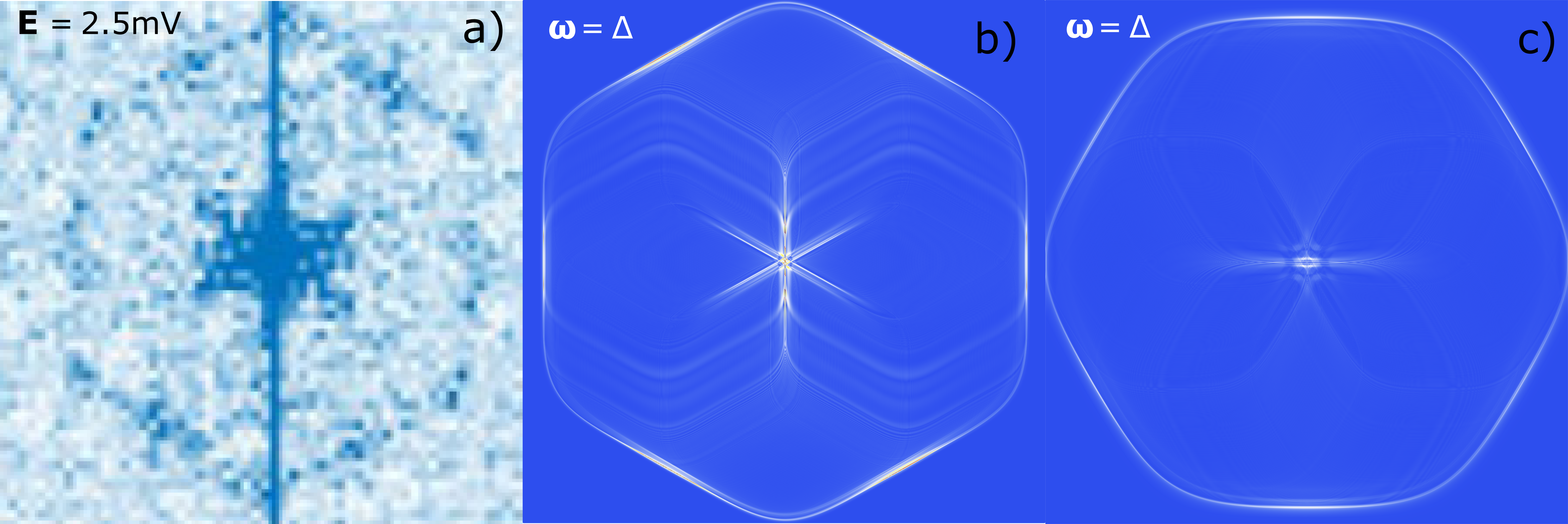}}
    \caption{Panel a): Experimental QPI at bias voltage $E=2.5meV$. Panel b): Theoretical QPI of the system with $\lambda_1\neq0$ and $\lambda_2=0$ . Panel c): Theoretical QPI of the system with two nonzero warping terms. At panel c) $\lambda_2>\lambda_1$. Orientation of the nematicity is $\Delta_{4y}$ for all cases.} 
	\label{Fig::FS_warping}
\end{figure}
\begin{figure*}
    \center{\includegraphics[width=0.8\linewidth]{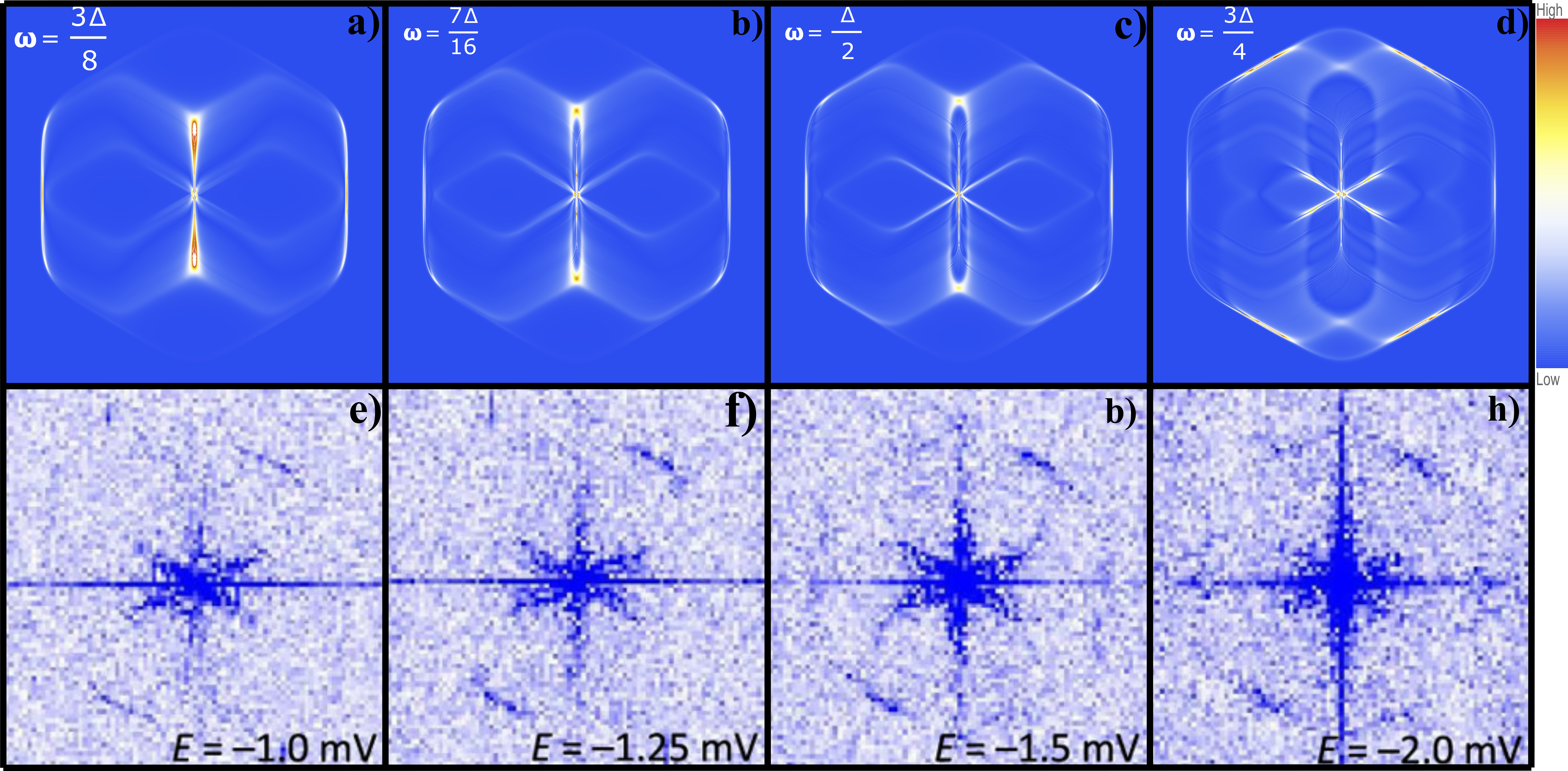}}
    \caption{Four top panels a)-d) shows QPI patterns at different bias voltages $\omega$ for system with only one warping $\lambda_1$. Four bottom panels e)-h) shows experimental results form the Ref.~\onlinecite{Chen2018} for $\Delta_{4y}$ orientation of the nematicity.} 
	\label{Fig::QPI_exp}
\end{figure*}
Recently, quasiparticle interference was measured in the thin film of Bi$_2$Te$_3$ placed on the iron-based superconductor FeTe$_{0.55}$Se$_{0.45}$ substrate~\cite{Chen2018}. The superconductivity was induced in the Bi$_2$Te$_3$ via proximity effect. The experimental QPI shows substrate induces a full-gap to the Bi$_2$Te$_3$ states with a magnitude between $0.5$meV and $1$meV. We show this experimental QPI in  Fig.~\ref{Fig::QPI_exp}e-h. At the bias voltage equal to $1$meV long wave vector reflexes appear, see Fig.~\ref{Fig::QPI_exp}e. These reflexes form two flat symmetrical stripes which are placed opposite to each other. Comparing the orientation of the measured QPI with the crystalline axis, authors of the experimental article find they observe nematic superconductivity with $\Delta_{4y}$ orientation.

It is questionable whether our Hamiltonian correctly describes properties of Bi$_2$Te$_3$/FeTe$_{0.55}$Se$_{0.45}$ heterostructure. Material FeTe$_{0.55}$Se$_{0.45}$ has a different symmetry group from the Bi$_2$Se$_3$, so we expect a different form a superconducting order parameter for such a structure.  Nevertheless, we can use our model to get the best fit for the experimental data. The first fitting `parameter' is the orientation of a QPI at high bias voltage. This orientation reproduces orientation of the isoenergy contour and depends on warping $\lambda_2/\lambda_1$ ratio. In our previous calculations we take this ratio equal to 2 taking it from DFT parameters. Such a choice gives a different from the experimental one orientation of the QPI. To fit theory to the experiment we set $\lambda_2=0$, keeping $\lambda_1\neq0$. Also, we make a small change to the another warping $\frac{\lambda_1m^2}{v^3}=0.17$, and chemical potential $\mu/m=2.0$. As we can see from Fig.~\ref{Fig::FS_warping} those new parameters give better fit for large bias $\omega \sim \Delta$.


We plot QPI patterns with new parameters at different bias voltages. Our results are shown in Fig.~\ref{Fig::QPI_exp}a-d. We add experimental QPI patterns from Ref.~\onlinecite{Chen2018} to compare with our theoretical results, see Fig.~\ref{Fig::QPI_exp}e-h. Based on our theory we provide mapping in bias between theory and experiment. In our model, the full gap is slightly smaller than $3/8\Delta$. We take the smallest bias voltage as $3/8\Delta$. Other voltages were taken to form the sequence $1:1.25:1.5:2$, as it was made in the experiment. In the experimental figures, the orientation of the axes is shown in Fig.~\ref{Fig::QPI_exp}h. Experimental axes are rotated by the angle $60^{\circ}$ anticlockwise in comparison with the axes we use in computations. 

At the lowest bias voltage experimental and theoretical patterns have only two flat long wave vector reflexes that correspond to $\mathbf{q}_1$ vector, see Figs.~\ref{Fig::QPI_exp}a,e. The same two reflexes dominate in Figs.~\ref{Fig::QPI_exp}b,f. In Fig.~\ref{Fig::QPI_exp}d other four sides of the hexagon appear. In Fig.~\ref{Fig::QPI_exp}d, QPI forms a perfect hexagon.

The experimental pattern has a well-noticeable six-pointed star in its center in all voltages besides the highest one, see Fig.~\ref{Fig::QPI_exp}e,f,g. Theoretical calculations predict a similar symmetrical star, except the lowest bias $\omega=3\Delta/8$. This star corresponds to the scattering vectors $\mathbf{q}_4$ and $\mathbf{q}_4'$. Note, that for the new parameters $\mathbf{q}_4'$ corresponds to the scattering across $y$ direction.

In contrast with the experimental picture, the theoretical has a strong reflex, associated with scattering $\mathbf{q}_2$, described in Sec.~\ref{sec_qpi} This reflex disappears at the higher voltages, see Fig.~\ref{Fig::QPI_exp}d,h. This indicates that induced superconductivity in Bi$_2$Te$_3$/FeTe$_{0.55}$Se$_{0.45}$ cannot be fully described by the native superconductivity in Bi$_2$Te$_3$ with the $E_u$ symmetry.

\section{Discussion}\label{sec_discussion}
 Without warpings $\lambda_1=\lambda_2=0$, QPI for different orientations of the nematicity can be obtained by the rotation at the relative angle of the nematicity $\alpha$. The presence of the warping breaks infinite rotational symmetry of the  Hamiltonian of the normal state $H_0$ down to three-fold symmetry~\cite{Fu2009} $C_{3v}$. This symmetry is incommensurate with the two-fold symmetry of the nematic order parameter. So, QPI images for the different orientations of the nematicity are different in presence of warping. These differences are visible in both coordinate and momentum spaces.

 Previously, only one warping $\lambda_1$ has been considered in the context of the nematic superconductivity~\cite{Fu2014}. Another warping $\lambda_2$ is equally important for the properties of the nematic superconductivity. Such a term determine the orientation of the Fermi surface for a normal state. If nematicity arises in the system


In addition, we compute QPI patterns for $\lambda_2=0$. In this case, we get QPI at high bias voltages similar to the experiment on the Bi$_2$Te$_3$/FeTe$_{0.55}$Se$_{0.45}$ heterostructure~\cite{Chen2018}.
In the nematic superconductors, nematicity manifests itself through the nematicity of the spectrum. So, we can expect, that some features of the QPI will be the same for different types of nematic superconductivity for the same material. We can see from Fig.~\ref{Fig::QPI_exp}, that $q_1$, $q_4$ and $q_{4'}$ scatterings are similar for both experimental and theoretical pictures. Such a similarity arises from the similarity of the energy spectrum of our Hamiltonian and experimental system. However, QPI depends not only on the properties of the energy spectra but also on the transition rate between the incident and scattered wavefunctions $\langle \psi_k|V|\psi_{k+q}\rangle$. These matrix elements generate selective rules for the possible scattering vectors. The absence of the $q_2$ vector in the experimental pictures implies that this scattering is forbidden by the selective rules. In the experiment, superconductivity is induced into the thin film from the bulk superconductor FeTe$_{0.55}$Se$_{0.45}$. Thus, induced superconductivity can have a different symmetry from the possible symmetries of the spontaneous superconductivity. These differences between the symmetries of the induced and spontaneous superconductivity may lead to different spin selective rules. We can conclude, that absence of the $q_2$ scattering in the experimental picture is a fingerprint that superconductivity induced from FeTe has a different symmetry from the $E_u$ symmetry of the spontaneous nematic superconductivity. 

Previously, QPI in nematic superconductors has been studied in Ref.~\onlinecite{Bao2018}. Study was performed for the 2D triangular tight-binding Hamiltonian. Obtained QPI differs significantly from our predictions and from the experimental results for Bi$_2$Te$_3$/FeTe$_{0.55}$Se$_{0.45}$ heterostructure.

Surface Andreev bound states can also contribute to the QPI along with the quasiparticle bulk states. Recent experimental studies have not found such surface states~\cite{Levy2013,Peng2013} while theoretical calculations predict surface Andreev bound states in superconducting topological insulators~\cite{Hsieh2012,Hao2015,Hao2017}. Effective Hamiltonian of the surface Andreev bound states have been obtained in Refs.~\onlinecite{Hsieh2012,Hao2017}. It appeared that these states are Majorana fermions. We checked that the matrix element that corresponds to the scattering of the surface states of the charged impurity vanishes $\langle \phi_i |V_0\tau_z | \phi_j \rangle =0$, where $\phi_i$ corresponds to the Majorana surface states. Thus, surface Andreev bound states have no contribution to the QPI in case of the scalar impurities.


In conclusion, we investigated details of QPI on the surface of the nematic superconductor induced by surface point-like scatter. Nematic behaviour of the QPI is visible if bias voltage is larger than the gap in the spectrum $\bar{\Delta}$ but smaller than the value of the order parameter $\bar{\Delta}<\omega<\Delta$. We showed that different orientations of the nematicity produce different QPI both in momentum and real spaces. We calculated QPI for a model with single warping $\lambda_2=0$ to match our results with the experimental QPI for Bi$_2$Te$_3$/FeTe$_{0.55}$Se$_{0.45}$ heterostructure. We found that long-wave and short-wave QPI pictures are similar. We pointed out that this similarity arises from the similarities of the spectra for the experimental structure and theoretical Hamiltonian. Absence of one of the scattering vectors in the experiment we attributed to the different symmetry of the experimental order parameter from the theoretical one.

\section*{Acknowledgment}

Authors acknowledge support by the Russian Scientific Foundation under Grant No 20-72-00030 and partial support from the Foundation for the Advancement of Theoretical Physics and Mathematics “BASIS”. 

\bibliographystyle{apsrevlong_no_issn_url}
\bibliography{nematic_QPI}

\begin{thebibliography}{57}
\expandafter\ifx\csname natexlab\endcsname\relax\def\natexlab#1{#1}\fi
\expandafter\ifx\csname bibnamefont\endcsname\relax
  \def\bibnamefont#1{#1}\fi
\expandafter\ifx\csname bibfnamefont\endcsname\relax
  \def\bibfnamefont#1{#1}\fi
\expandafter\ifx\csname citenamefont\endcsname\relax
  \def\citenamefont#1{#1}\fi

\bibitem[{\citenamefont{Sasaki et~al.}(2011)\citenamefont{Sasaki, Kriener,
  Segawa, Yada, Tanaka, Sato, and Ando}}]{Sasaki2011}
\bibinfo{author}{\bibfnamefont{S.}~\bibnamefont{Sasaki}},
  \bibinfo{author}{\bibfnamefont{M.}~\bibnamefont{Kriener}},
  \bibinfo{author}{\bibfnamefont{K.}~\bibnamefont{Segawa}},
  \bibinfo{author}{\bibfnamefont{K.}~\bibnamefont{Yada}},
  \bibinfo{author}{\bibfnamefont{Y.}~\bibnamefont{Tanaka}},
  \bibinfo{author}{\bibfnamefont{M.}~\bibnamefont{Sato}}, \bibnamefont{and}
  \bibinfo{author}{\bibfnamefont{Y.}~\bibnamefont{Ando}},
  {``}\bibinfo{title}{Topological Superconductivity in
  ${\mathrm{Cu}}_{x}{\mathrm{Bi}}_{2}{\mathrm{Se}}_{3}$},{''}
  \bibinfo{journal}{Phys. Rev. Lett.} \textbf{\bibinfo{volume}{107}},
  \bibinfo{pages}{217001} (\bibinfo{year}{2011}).

\bibitem[{\citenamefont{Chen et~al.}(2018)\citenamefont{Chen, Chen, Yang, Du,
  and Wen}}]{Chen2018}
\bibinfo{author}{\bibfnamefont{M.}~\bibnamefont{Chen}},
  \bibinfo{author}{\bibfnamefont{X.}~\bibnamefont{Chen}},
  \bibinfo{author}{\bibfnamefont{H.}~\bibnamefont{Yang}},
  \bibinfo{author}{\bibfnamefont{Z.}~\bibnamefont{Du}}, \bibnamefont{and}
  \bibinfo{author}{\bibfnamefont{H.-H.} \bibnamefont{Wen}},
  {``}\bibinfo{title}{Superconductivity with twofold symmetry in
  Bi2Te3/{FeTe}0.55Se0.45 heterostructures},{''} \bibinfo{journal}{Science
  Advances} \textbf{\bibinfo{volume}{4}}, \bibinfo{pages}{eaat1084}
  (\bibinfo{year}{2018}).

\bibitem[{\citenamefont{Charpentier et~al.}(2017)\citenamefont{Charpentier,
  Galletti, Kunakova, Arpaia, Song, Baghdadi, Wang, Kalaboukhov, Olsson, Tafuri
  et~al.}}]{Charpentier2017}
\bibinfo{author}{\bibfnamefont{S.}~\bibnamefont{Charpentier}},
  \bibinfo{author}{\bibfnamefont{L.}~\bibnamefont{Galletti}},
  \bibinfo{author}{\bibfnamefont{G.}~\bibnamefont{Kunakova}},
  \bibinfo{author}{\bibfnamefont{R.}~\bibnamefont{Arpaia}},
  \bibinfo{author}{\bibfnamefont{Y.}~\bibnamefont{Song}},
  \bibinfo{author}{\bibfnamefont{R.}~\bibnamefont{Baghdadi}},
  \bibinfo{author}{\bibfnamefont{S.~M.} \bibnamefont{Wang}},
  \bibinfo{author}{\bibfnamefont{A.}~\bibnamefont{Kalaboukhov}},
  \bibinfo{author}{\bibfnamefont{E.}~\bibnamefont{Olsson}},
  \bibinfo{author}{\bibfnamefont{F.}~\bibnamefont{Tafuri}},
  \bibnamefont{et~al.}, {``}\bibinfo{title}{Induced unconventional
  superconductivity on the surface states of Bi2Te3 topological insulator},{''}
  \bibinfo{journal}{Nature Communications} \textbf{\bibinfo{volume}{8}},
  \bibinfo{pages}{2019} (\bibinfo{year}{2017}).

\bibitem[{\citenamefont{Du et~al.}(2017)\citenamefont{Du, Shao, Yang, Du, Fang,
  Wang, Ran, Wen, Zhang, Yang et~al.}}]{Du2017}
\bibinfo{author}{\bibfnamefont{G.}~\bibnamefont{Du}},
  \bibinfo{author}{\bibfnamefont{J.}~\bibnamefont{Shao}},
  \bibinfo{author}{\bibfnamefont{X.}~\bibnamefont{Yang}},
  \bibinfo{author}{\bibfnamefont{Z.}~\bibnamefont{Du}},
  \bibinfo{author}{\bibfnamefont{D.}~\bibnamefont{Fang}},
  \bibinfo{author}{\bibfnamefont{J.}~\bibnamefont{Wang}},
  \bibinfo{author}{\bibfnamefont{K.}~\bibnamefont{Ran}},
  \bibinfo{author}{\bibfnamefont{J.}~\bibnamefont{Wen}},
  \bibinfo{author}{\bibfnamefont{C.}~\bibnamefont{Zhang}},
  \bibinfo{author}{\bibfnamefont{H.}~\bibnamefont{Yang}}, \bibnamefont{et~al.},
  {``}\bibinfo{title}{Drive the Dirac electrons into Cooper pairs in
  {SrxBi}2Se3},{''} \bibinfo{journal}{Nature Communications}
  \textbf{\bibinfo{volume}{8}} (\bibinfo{year}{2017}).

\bibitem[{\citenamefont{Yonezawa et~al.}(2016)\citenamefont{Yonezawa, Tajiri,
  Nakata, Nagai, Wang, Segawa, Ando, and Maeno}}]{Yonezawa2016}
\bibinfo{author}{\bibfnamefont{S.}~\bibnamefont{Yonezawa}},
  \bibinfo{author}{\bibfnamefont{K.}~\bibnamefont{Tajiri}},
  \bibinfo{author}{\bibfnamefont{S.}~\bibnamefont{Nakata}},
  \bibinfo{author}{\bibfnamefont{Y.}~\bibnamefont{Nagai}},
  \bibinfo{author}{\bibfnamefont{Z.}~\bibnamefont{Wang}},
  \bibinfo{author}{\bibfnamefont{K.}~\bibnamefont{Segawa}},
  \bibinfo{author}{\bibfnamefont{Y.}~\bibnamefont{Ando}}, \bibnamefont{and}
  \bibinfo{author}{\bibfnamefont{Y.}~\bibnamefont{Maeno}},
  {``}\bibinfo{title}{Thermodynamic evidence for nematic superconductivity in
  {CuxBi}2Se3},{''} \bibinfo{journal}{Nature Physics}
  \textbf{\bibinfo{volume}{13}}, \bibinfo{pages}{123} (\bibinfo{year}{2016}).

\bibitem[{\citenamefont{Kasahara et~al.}(2012)\citenamefont{Kasahara, Shi,
  Hashimoto, Tonegawa, Mizukami, Shibauchi, Sugimoto, Fukuda, Terashima,
  Nevidomskyy et~al.}}]{Kasahara2012}
\bibinfo{author}{\bibfnamefont{S.}~\bibnamefont{Kasahara}},
  \bibinfo{author}{\bibfnamefont{H.~J.} \bibnamefont{Shi}},
  \bibinfo{author}{\bibfnamefont{K.}~\bibnamefont{Hashimoto}},
  \bibinfo{author}{\bibfnamefont{S.}~\bibnamefont{Tonegawa}},
  \bibinfo{author}{\bibfnamefont{Y.}~\bibnamefont{Mizukami}},
  \bibinfo{author}{\bibfnamefont{T.}~\bibnamefont{Shibauchi}},
  \bibinfo{author}{\bibfnamefont{K.}~\bibnamefont{Sugimoto}},
  \bibinfo{author}{\bibfnamefont{T.}~\bibnamefont{Fukuda}},
  \bibinfo{author}{\bibfnamefont{T.}~\bibnamefont{Terashima}},
  \bibinfo{author}{\bibfnamefont{A.~H.} \bibnamefont{Nevidomskyy}},
  \bibnamefont{et~al.}, {``}\bibinfo{title}{Electronic nematicity above the
  structural and superconducting transition in {BaFe}2(As1-{xP} x )2},{''}
  \bibinfo{journal}{Nature} \textbf{\bibinfo{volume}{486}},
  \bibinfo{pages}{382} (\bibinfo{year}{2012}).

\bibitem[{\citenamefont{Asaba et~al.}(2017)\citenamefont{Asaba, Lawson,
  Tinsman, Chen, Corbae, Li, Qiu, Hor, Fu, and Li}}]{Asaba2017}
\bibinfo{author}{\bibfnamefont{T.}~\bibnamefont{Asaba}},
  \bibinfo{author}{\bibfnamefont{B.~J.} \bibnamefont{Lawson}},
  \bibinfo{author}{\bibfnamefont{C.}~\bibnamefont{Tinsman}},
  \bibinfo{author}{\bibfnamefont{L.}~\bibnamefont{Chen}},
  \bibinfo{author}{\bibfnamefont{P.}~\bibnamefont{Corbae}},
  \bibinfo{author}{\bibfnamefont{G.}~\bibnamefont{Li}},
  \bibinfo{author}{\bibfnamefont{Y.}~\bibnamefont{Qiu}},
  \bibinfo{author}{\bibfnamefont{Y.~S.} \bibnamefont{Hor}},
  \bibinfo{author}{\bibfnamefont{L.}~\bibnamefont{Fu}}, \bibnamefont{and}
  \bibinfo{author}{\bibfnamefont{L.}~\bibnamefont{Li}},
  {``}\bibinfo{title}{Rotational Symmetry Breaking in a Trigonal Superconductor
  Nb-doped ${\mathrm{Bi}}_{2}{\mathrm{Se}}_{3}$},{''} \bibinfo{journal}{Phys.
  Rev. X} \textbf{\bibinfo{volume}{7}}, \bibinfo{pages}{011009}
  (\bibinfo{year}{2017}).

\bibitem[{\citenamefont{Li et~al.}(2018)\citenamefont{Li, Wang, Zhang, Feng,
  Jiang, Han, Chen, Ye, Gao, Jia et~al.}}]{Li2018}
\bibinfo{author}{\bibfnamefont{Z.}~\bibnamefont{Li}},
  \bibinfo{author}{\bibfnamefont{M.}~\bibnamefont{Wang}},
  \bibinfo{author}{\bibfnamefont{D.}~\bibnamefont{Zhang}},
  \bibinfo{author}{\bibfnamefont{N.}~\bibnamefont{Feng}},
  \bibinfo{author}{\bibfnamefont{W.}~\bibnamefont{Jiang}},
  \bibinfo{author}{\bibfnamefont{C.}~\bibnamefont{Han}},
  \bibinfo{author}{\bibfnamefont{W.}~\bibnamefont{Chen}},
  \bibinfo{author}{\bibfnamefont{M.}~\bibnamefont{Ye}},
  \bibinfo{author}{\bibfnamefont{C.}~\bibnamefont{Gao}},
  \bibinfo{author}{\bibfnamefont{J.}~\bibnamefont{Jia}}, \bibnamefont{et~al.},
  {``}\bibinfo{title}{Possible structural origin of superconductivity in
  Sr-doped $\mathrm{B}{\mathrm{i}}_{2}\mathrm{S}{\mathrm{e}}_{3}$},{''}
  \bibinfo{journal}{Phys. Rev. Materials} \textbf{\bibinfo{volume}{2}},
  \bibinfo{pages}{014201} (\bibinfo{year}{2018}).

\bibitem[{\citenamefont{Venderbos et~al.}(2018)\citenamefont{Venderbos, Savary,
  Ruhman, Lee, and Fu}}]{Venderbos2018}
\bibinfo{author}{\bibfnamefont{J.~W.~F.} \bibnamefont{Venderbos}},
  \bibinfo{author}{\bibfnamefont{L.}~\bibnamefont{Savary}},
  \bibinfo{author}{\bibfnamefont{J.}~\bibnamefont{Ruhman}},
  \bibinfo{author}{\bibfnamefont{P.~A.} \bibnamefont{Lee}}, \bibnamefont{and}
  \bibinfo{author}{\bibfnamefont{L.}~\bibnamefont{Fu}},
  {``}\bibinfo{title}{Pairing States of Spin-$\frac{3}{2}$ Fermions:
  Symmetry-Enforced Topological Gap Functions},{''} \bibinfo{journal}{Phys.
  Rev. X} \textbf{\bibinfo{volume}{8}}, \bibinfo{pages}{011029}
  (\bibinfo{year}{2018}).

\bibitem[{\citenamefont{Brydon et~al.}(2014)\citenamefont{Brydon, Das~Sarma,
  Hui, and Sau}}]{Brydon2014}
\bibinfo{author}{\bibfnamefont{P.~M.~R.} \bibnamefont{Brydon}},
  \bibinfo{author}{\bibfnamefont{S.}~\bibnamefont{Das~Sarma}},
  \bibinfo{author}{\bibfnamefont{H.-Y.} \bibnamefont{Hui}}, \bibnamefont{and}
  \bibinfo{author}{\bibfnamefont{J.~D.} \bibnamefont{Sau}},
  {``}\bibinfo{title}{Odd-parity superconductivity from phonon-mediated
  pairing: Application to
  ${\mathrm{Cu}}_{x}{\mathrm{Bi}}_{2}{\mathrm{Se}}_{3}$},{''}
  \bibinfo{journal}{Phys. Rev. B} \textbf{\bibinfo{volume}{90}},
  \bibinfo{pages}{184512} (\bibinfo{year}{2014}).

\bibitem[{\citenamefont{Hecker and Schmalian}(2017)}]{Hecker2018}
\bibinfo{author}{\bibfnamefont{M.}~\bibnamefont{Hecker}} \bibnamefont{and}
  \bibinfo{author}{\bibfnamefont{J.}~\bibnamefont{Schmalian}},
  {``}\bibinfo{title}{Vestigial nematic order and superconductivity in the
  doped topological insulator Cu$_x$Bi$_2$Se$_3$},{''} \bibinfo{journal}{npj
  Quantum Mater.} p.~\bibinfo{pages}{26} (\bibinfo{year}{2017}).

\bibitem[{\citenamefont{Chiba et~al.}(2017)\citenamefont{Chiba, Takahashi, and
  Bauer}}]{Chiba2017}
\bibinfo{author}{\bibfnamefont{T.}~\bibnamefont{Chiba}},
  \bibinfo{author}{\bibfnamefont{S.}~\bibnamefont{Takahashi}},
  \bibnamefont{and} \bibinfo{author}{\bibfnamefont{G.~E.~W.}
  \bibnamefont{Bauer}}, {``}\bibinfo{title}{Magnetic-proximity-induced
  magnetoresistance on topological insulators},{''} \bibinfo{journal}{Phys.
  Rev. B} \textbf{\bibinfo{volume}{95}}, \bibinfo{pages}{094428}
  (\bibinfo{year}{2017}).

\bibitem[{\citenamefont{Hor et~al.}(2010)\citenamefont{Hor, Williams,
  Checkelsky, Roushan, Seo, Xu, Zandbergen, Yazdani, Ong, and Cava}}]{Hor2010}
\bibinfo{author}{\bibfnamefont{Y.~S.} \bibnamefont{Hor}},
  \bibinfo{author}{\bibfnamefont{A.~J.} \bibnamefont{Williams}},
  \bibinfo{author}{\bibfnamefont{J.~G.} \bibnamefont{Checkelsky}},
  \bibinfo{author}{\bibfnamefont{P.}~\bibnamefont{Roushan}},
  \bibinfo{author}{\bibfnamefont{J.}~\bibnamefont{Seo}},
  \bibinfo{author}{\bibfnamefont{Q.}~\bibnamefont{Xu}},
  \bibinfo{author}{\bibfnamefont{H.~W.} \bibnamefont{Zandbergen}},
  \bibinfo{author}{\bibfnamefont{A.}~\bibnamefont{Yazdani}},
  \bibinfo{author}{\bibfnamefont{N.~P.} \bibnamefont{Ong}}, \bibnamefont{and}
  \bibinfo{author}{\bibfnamefont{R.~J.} \bibnamefont{Cava}},
  {``}\bibinfo{title}{Superconductivity in
  ${\mathrm{Cu}}_{x}{\mathrm{Bi}}_{2}{\mathrm{Se}}_{3}$ and its Implications
  for Pairing in the Undoped Topological Insulator},{''}
  \bibinfo{journal}{Phys. Rev. Lett.} \textbf{\bibinfo{volume}{104}},
  \bibinfo{pages}{057001} (\bibinfo{year}{2010}).

\bibitem[{\citenamefont{Kirzhner et~al.}(2012)\citenamefont{Kirzhner, Lahoud,
  Chaska, Salman, and Kanigel}}]{Kirzhner2012}
\bibinfo{author}{\bibfnamefont{T.}~\bibnamefont{Kirzhner}},
  \bibinfo{author}{\bibfnamefont{E.}~\bibnamefont{Lahoud}},
  \bibinfo{author}{\bibfnamefont{K.~B.} \bibnamefont{Chaska}},
  \bibinfo{author}{\bibfnamefont{Z.}~\bibnamefont{Salman}}, \bibnamefont{and}
  \bibinfo{author}{\bibfnamefont{A.}~\bibnamefont{Kanigel}},
  {``}\bibinfo{title}{Point-contact spectroscopy of
  Cu${}_{0.2}$Bi${}_{2}$Se${}_{3}$ single crystals},{''}
  \bibinfo{journal}{Phys. Rev. B} \textbf{\bibinfo{volume}{86}},
  \bibinfo{pages}{064517} (\bibinfo{year}{2012}).

\bibitem[{\citenamefont{Kriener et~al.}(2011)\citenamefont{Kriener, Segawa,
  Ren, Sasaki, and Ando}}]{Kriener2011}
\bibinfo{author}{\bibfnamefont{M.}~\bibnamefont{Kriener}},
  \bibinfo{author}{\bibfnamefont{K.}~\bibnamefont{Segawa}},
  \bibinfo{author}{\bibfnamefont{Z.}~\bibnamefont{Ren}},
  \bibinfo{author}{\bibfnamefont{S.}~\bibnamefont{Sasaki}}, \bibnamefont{and}
  \bibinfo{author}{\bibfnamefont{Y.}~\bibnamefont{Ando}},
  {``}\bibinfo{title}{Bulk Superconducting Phase with a Full Energy Gap in the
  Doped Topological Insulator
  ${\mathrm{Cu}}_{x}{\mathrm{Bi}}_{2}{\mathrm{Se}}_{3}$},{''}
  \bibinfo{journal}{Phys. Rev. Lett.} \textbf{\bibinfo{volume}{106}},
  \bibinfo{pages}{127004} (\bibinfo{year}{2011}).

\bibitem[{\citenamefont{Kuntsevich et~al.}(2018)\citenamefont{Kuntsevich,
  Bryzgalov, Prudkoglyad, Martovitskii, Selivanov, and
  Chizhevskii}}]{Kuntsevich2018}
\bibinfo{author}{\bibfnamefont{A.~Y.} \bibnamefont{Kuntsevich}},
  \bibinfo{author}{\bibfnamefont{M.~A.} \bibnamefont{Bryzgalov}},
  \bibinfo{author}{\bibfnamefont{V.~A.} \bibnamefont{Prudkoglyad}},
  \bibinfo{author}{\bibfnamefont{V.~P.} \bibnamefont{Martovitskii}},
  \bibinfo{author}{\bibfnamefont{Y.~G.} \bibnamefont{Selivanov}},
  \bibnamefont{and} \bibinfo{author}{\bibfnamefont{E.~G.}
  \bibnamefont{Chizhevskii}}, {``}\bibinfo{title}{Structural distortion behind
  the nematic superconductivity in Sr x Bi2Se3},{''},
  \textbf{\bibinfo{volume}{20}}, \bibinfo{pages}{103022}
  (\bibinfo{year}{2018}).

\bibitem[{\citenamefont{Kuntsevich et~al.}(2019)\citenamefont{Kuntsevich,
  Bryzgalov, Akzyanov, Martovitskii, Rakhmanov, and
  Selivanov}}]{Kuntsevich2019}
\bibinfo{author}{\bibfnamefont{A.~Y.} \bibnamefont{Kuntsevich}},
  \bibinfo{author}{\bibfnamefont{M.~A.} \bibnamefont{Bryzgalov}},
  \bibinfo{author}{\bibfnamefont{R.~S.} \bibnamefont{Akzyanov}},
  \bibinfo{author}{\bibfnamefont{V.~P.} \bibnamefont{Martovitskii}},
  \bibinfo{author}{\bibfnamefont{A.~L.} \bibnamefont{Rakhmanov}},
  \bibnamefont{and} \bibinfo{author}{\bibfnamefont{Y.~G.}
  \bibnamefont{Selivanov}}, {``}\bibinfo{title}{Strain-driven nematicity of
  odd-parity superconductivity in
  ${\mathrm{Sr}}_{x}{\mathrm{Bi}}_{2}{\mathrm{Se}}_{3}$},{''}
  \bibinfo{journal}{Phys. Rev. B} \textbf{\bibinfo{volume}{100}},
  \bibinfo{pages}{224509} (\bibinfo{year}{2019}).

\bibitem[{\citenamefont{Kozii and Fu}(2015)}]{Kozii2015}
\bibinfo{author}{\bibfnamefont{V.}~\bibnamefont{Kozii}} \bibnamefont{and}
  \bibinfo{author}{\bibfnamefont{L.}~\bibnamefont{Fu}},
  {``}\bibinfo{title}{Odd-Parity Superconductivity in the Vicinity of Inversion
  Symmetry Breaking in Spin-Orbit-Coupled Systems},{''} \bibinfo{journal}{Phys.
  Rev. Lett.} \textbf{\bibinfo{volume}{115}}, \bibinfo{pages}{207002}
  (\bibinfo{year}{2015}).

\bibitem[{\citenamefont{Matano et~al.}(2016)\citenamefont{Matano, Kriener,
  Segawa, Ando, and qing Zheng}}]{Matano2016}
\bibinfo{author}{\bibfnamefont{K.}~\bibnamefont{Matano}},
  \bibinfo{author}{\bibfnamefont{M.}~\bibnamefont{Kriener}},
  \bibinfo{author}{\bibfnamefont{K.}~\bibnamefont{Segawa}},
  \bibinfo{author}{\bibfnamefont{Y.}~\bibnamefont{Ando}}, \bibnamefont{and}
  \bibinfo{author}{\bibfnamefont{G.}~\bibnamefont{qing Zheng}},
  {``}\bibinfo{title}{Spin-rotation symmetry breaking in the superconducting
  state of {CuxBi}2Se3},{''} \bibinfo{journal}{Nature Physics}
  \textbf{\bibinfo{volume}{12}}, \bibinfo{pages}{852} (\bibinfo{year}{2016}).

\bibitem[{\citenamefont{Tao et~al.}(2018)\citenamefont{Tao, Yan, Liu, Wang,
  Ando, Wang, Zhang, and Feng}}]{Tao2018}
\bibinfo{author}{\bibfnamefont{R.}~\bibnamefont{Tao}},
  \bibinfo{author}{\bibfnamefont{Y.-J.} \bibnamefont{Yan}},
  \bibinfo{author}{\bibfnamefont{X.}~\bibnamefont{Liu}},
  \bibinfo{author}{\bibfnamefont{Z.-W.} \bibnamefont{Wang}},
  \bibinfo{author}{\bibfnamefont{Y.}~\bibnamefont{Ando}},
  \bibinfo{author}{\bibfnamefont{Q.-H.} \bibnamefont{Wang}},
  \bibinfo{author}{\bibfnamefont{T.}~\bibnamefont{Zhang}}, \bibnamefont{and}
  \bibinfo{author}{\bibfnamefont{D.-L.} \bibnamefont{Feng}},
  {``}\bibinfo{title}{Direct Visualization of the Nematic Superconductivity in
  ${\mathrm{Cu}}_{x}{\mathrm{Bi}}_{2}{\mathrm{Se}}_{3}$},{''}
  \bibinfo{journal}{Phys. Rev. X} \textbf{\bibinfo{volume}{8}},
  \bibinfo{pages}{041024} (\bibinfo{year}{2018}).

\bibitem[{\citenamefont{Sirohi et~al.}(2018)\citenamefont{Sirohi, Das, Neha,
  Jat, Patnaik, and Sheet}}]{Sirohi2018}
\bibinfo{author}{\bibfnamefont{A.}~\bibnamefont{Sirohi}},
  \bibinfo{author}{\bibfnamefont{S.}~\bibnamefont{Das}},
  \bibinfo{author}{\bibfnamefont{P.}~\bibnamefont{Neha}},
  \bibinfo{author}{\bibfnamefont{K.~S.} \bibnamefont{Jat}},
  \bibinfo{author}{\bibfnamefont{S.}~\bibnamefont{Patnaik}}, \bibnamefont{and}
  \bibinfo{author}{\bibfnamefont{G.}~\bibnamefont{Sheet}},
  {``}\bibinfo{title}{Low-energy excitations and non-BCS superconductivity in
  ${\mathrm{Nb}}_{x}\text{\ensuremath{-}}{\mathrm{Bi}}_{2}{\mathrm{Se}}_{3}$},{''}
  \bibinfo{journal}{Phys. Rev. B} \textbf{\bibinfo{volume}{98}},
  \bibinfo{pages}{094523} (\bibinfo{year}{2018}).

\bibitem[{\citenamefont{Pan et~al.}(2016)\citenamefont{Pan, Nikitin, Araizi,
  Huang, Matsushita, Naka, and de~Visser}}]{Pan2016}
\bibinfo{author}{\bibfnamefont{Y.}~\bibnamefont{Pan}},
  \bibinfo{author}{\bibfnamefont{A.~M.} \bibnamefont{Nikitin}},
  \bibinfo{author}{\bibfnamefont{G.~K.} \bibnamefont{Araizi}},
  \bibinfo{author}{\bibfnamefont{Y.~K.} \bibnamefont{Huang}},
  \bibinfo{author}{\bibfnamefont{Y.}~\bibnamefont{Matsushita}},
  \bibinfo{author}{\bibfnamefont{T.}~\bibnamefont{Naka}}, \bibnamefont{and}
  \bibinfo{author}{\bibfnamefont{A.}~\bibnamefont{de~Visser}},
  {``}\bibinfo{title}{Rotational symmetry breaking in the topological
  superconductor SrxBi2Se3 probed by upper-critical field experiments},{''}
  \bibinfo{journal}{Scientific Reports} \textbf{\bibinfo{volume}{6}},
  \bibinfo{pages}{28632} (\bibinfo{year}{2016}).

\bibitem[{\citenamefont{Fu and Berg}(2010)}]{Fu2010}
\bibinfo{author}{\bibfnamefont{L.}~\bibnamefont{Fu}} \bibnamefont{and}
  \bibinfo{author}{\bibfnamefont{E.}~\bibnamefont{Berg}},
  {``}\bibinfo{title}{Odd-Parity Topological Superconductors: Theory and
  Application to ${\mathrm{Cu}}_{x}{\mathrm{Bi}}_{2}{\mathrm{Se}}_{3}$},{''}
  \bibinfo{journal}{Phys. Rev. Lett.} \textbf{\bibinfo{volume}{105}},
  \bibinfo{pages}{097001} (\bibinfo{year}{2010}).

\bibitem[{\citenamefont{Fu}(2014)}]{Fu2014}
\bibinfo{author}{\bibfnamefont{L.}~\bibnamefont{Fu}},
  {``}\bibinfo{title}{Odd-parity topological superconductor with nematic order:
  Application to ${\mathrm{Cu}}_{x}{\mathrm{Bi}}_{2}{\mathrm{Se}}_{3}$},{''}
  \bibinfo{journal}{Phys. Rev. B} \textbf{\bibinfo{volume}{90}},
  \bibinfo{pages}{100509} (\bibinfo{year}{2014}).

\bibitem[{\citenamefont{Venderbos et~al.}(2016)\citenamefont{Venderbos, Kozii,
  and Fu}}]{Venderbos2016}
\bibinfo{author}{\bibfnamefont{J.~W.~F.} \bibnamefont{Venderbos}},
  \bibinfo{author}{\bibfnamefont{V.}~\bibnamefont{Kozii}}, \bibnamefont{and}
  \bibinfo{author}{\bibfnamefont{L.}~\bibnamefont{Fu}},
  {``}\bibinfo{title}{Identification of nematic superconductivity from the
  upper critical field},{''} \bibinfo{journal}{Phys. Rev. B}
  \textbf{\bibinfo{volume}{94}}, \bibinfo{pages}{094522}
  (\bibinfo{year}{2016}).

\bibitem[{\citenamefont{Kawai et~al.}(2020)\citenamefont{Kawai, Wang, Kandori,
  Honoki, Matano, Kambe, and qing Zheng}}]{Kawai2020}
\bibinfo{author}{\bibfnamefont{T.}~\bibnamefont{Kawai}},
  \bibinfo{author}{\bibfnamefont{C.~G.} \bibnamefont{Wang}},
  \bibinfo{author}{\bibfnamefont{Y.}~\bibnamefont{Kandori}},
  \bibinfo{author}{\bibfnamefont{Y.}~\bibnamefont{Honoki}},
  \bibinfo{author}{\bibfnamefont{K.}~\bibnamefont{Matano}},
  \bibinfo{author}{\bibfnamefont{T.}~\bibnamefont{Kambe}}, \bibnamefont{and}
  \bibinfo{author}{\bibfnamefont{G.}~\bibnamefont{qing Zheng}},
  {``}\bibinfo{title}{Direction and symmetry transition of the vector order
  parameter in topological superconductors {CuxBi}2Se3},{''}
  \bibinfo{journal}{Nature Communications} \textbf{\bibinfo{volume}{11}}
  (\bibinfo{year}{2020}).

\bibitem[{\citenamefont{Andersen et~al.}(2018)\citenamefont{Andersen, Wang,
  Lorenz, and Ando}}]{Andersen2018}
\bibinfo{author}{\bibfnamefont{L.}~\bibnamefont{Andersen}},
  \bibinfo{author}{\bibfnamefont{Z.}~\bibnamefont{Wang}},
  \bibinfo{author}{\bibfnamefont{T.}~\bibnamefont{Lorenz}}, \bibnamefont{and}
  \bibinfo{author}{\bibfnamefont{Y.}~\bibnamefont{Ando}},
  {``}\bibinfo{title}{Nematic superconductivity in
  ${\mathrm{Cu}}_{1.5}{(\mathrm{PbSe})}_{5}{({\mathrm{Bi}}_{2}{\mathrm{Se}}_{3})}_{6}$},{''}
  \bibinfo{journal}{Phys. Rev. B} \textbf{\bibinfo{volume}{98}},
  \bibinfo{pages}{220512} (\bibinfo{year}{2018}).

\bibitem[{\citenamefont{Yonezawa}(2018)}]{Yonezawa2018}
\bibinfo{author}{\bibfnamefont{S.}~\bibnamefont{Yonezawa}},
  {``}\bibinfo{title}{Nematic Superconductivity in Doped Bi2Se3 Topological
  Superconductors},{''} \bibinfo{journal}{Condensed Matter}
  \textbf{\bibinfo{volume}{4}}, \bibinfo{pages}{2} (\bibinfo{year}{2018}).

\bibitem[{\citenamefont{Kostylev et~al.}(2020)\citenamefont{Kostylev, Yonezawa,
  Wang, Ando, and Maeno}}]{Kostylev2020}
\bibinfo{author}{\bibfnamefont{I.}~\bibnamefont{Kostylev}},
  \bibinfo{author}{\bibfnamefont{S.}~\bibnamefont{Yonezawa}},
  \bibinfo{author}{\bibfnamefont{Z.}~\bibnamefont{Wang}},
  \bibinfo{author}{\bibfnamefont{Y.}~\bibnamefont{Ando}}, \bibnamefont{and}
  \bibinfo{author}{\bibfnamefont{Y.}~\bibnamefont{Maeno}},
  {``}\bibinfo{title}{Uniaxial-strain control of nematic superconductivity in
  SrxBi2Se3},{''} \bibinfo{journal}{Nature Communications}
  \textbf{\bibinfo{volume}{11}}, \bibinfo{pages}{4152} (\bibinfo{year}{2020}).

\bibitem[{\citenamefont{Kuroda et~al.}(2010)\citenamefont{Kuroda, Arita,
  Miyamoto, Ye, Jiang, Kimura, Krasovskii, Chulkov, Iwasawa, Okuda
  et~al.}}]{Kuroda2010}
\bibinfo{author}{\bibfnamefont{K.}~\bibnamefont{Kuroda}},
  \bibinfo{author}{\bibfnamefont{M.}~\bibnamefont{Arita}},
  \bibinfo{author}{\bibfnamefont{K.}~\bibnamefont{Miyamoto}},
  \bibinfo{author}{\bibfnamefont{M.}~\bibnamefont{Ye}},
  \bibinfo{author}{\bibfnamefont{J.}~\bibnamefont{Jiang}},
  \bibinfo{author}{\bibfnamefont{A.}~\bibnamefont{Kimura}},
  \bibinfo{author}{\bibfnamefont{E.~E.} \bibnamefont{Krasovskii}},
  \bibinfo{author}{\bibfnamefont{E.~V.} \bibnamefont{Chulkov}},
  \bibinfo{author}{\bibfnamefont{H.}~\bibnamefont{Iwasawa}},
  \bibinfo{author}{\bibfnamefont{T.}~\bibnamefont{Okuda}},
  \bibnamefont{et~al.}, {``}\bibinfo{title}{Hexagonally Deformed Fermi Surface
  of the 3D Topological Insulator ${\mathrm{Bi}}_{2}{\mathrm{Se}}_{3}$},{''}
  \bibinfo{journal}{Phys. Rev. Lett.} \textbf{\bibinfo{volume}{105}},
  \bibinfo{pages}{076802} (\bibinfo{year}{2010}).

\bibitem[{\citenamefont{Fu}(2009)}]{Fu2009}
\bibinfo{author}{\bibfnamefont{L.}~\bibnamefont{Fu}},
  {``}\bibinfo{title}{Hexagonal Warping Effects in the Surface States of the
  Topological Insulator ${\mathrm{Bi}}_{2}{\mathrm{Te}}_{3}$},{''}
  \bibinfo{journal}{Phys. Rev. Lett.} \textbf{\bibinfo{volume}{103}},
  \bibinfo{pages}{266801} (\bibinfo{year}{2009}).

\bibitem[{\citenamefont{Akzyanov and Rakhmanov}(2018)}]{Akzyanov2018}
\bibinfo{author}{\bibfnamefont{R.~S.} \bibnamefont{Akzyanov}} \bibnamefont{and}
  \bibinfo{author}{\bibfnamefont{A.~L.} \bibnamefont{Rakhmanov}},
  {``}\bibinfo{title}{Surface charge conductivity of a topological insulator in
  a magnetic field: The effect of hexagonal warping},{''}
  \bibinfo{journal}{Phys. Rev. B} \textbf{\bibinfo{volume}{97}},
  \bibinfo{pages}{075421} (\bibinfo{year}{2018}).

\bibitem[{\citenamefont{Akzyanov and Rakhmanov}(2019)}]{Akzyanov2019}
\bibinfo{author}{\bibfnamefont{R.~S.} \bibnamefont{Akzyanov}} \bibnamefont{and}
  \bibinfo{author}{\bibfnamefont{A.~L.} \bibnamefont{Rakhmanov}},
  {``}\bibinfo{title}{Bulk and surface spin conductivity in topological
  insulators with hexagonal warping},{''} \bibinfo{journal}{Phys. Rev. B}
  \textbf{\bibinfo{volume}{99}}, \bibinfo{pages}{045436}
  (\bibinfo{year}{2019}).

\bibitem[{\citenamefont{Avraham et~al.}(2018)\citenamefont{Avraham, Reiner,
  Kumar-Nayak, Morali, Batabyal, Yan, and Beidenkopf}}]{Avraham2018}
\bibinfo{author}{\bibfnamefont{N.}~\bibnamefont{Avraham}},
  \bibinfo{author}{\bibfnamefont{J.}~\bibnamefont{Reiner}},
  \bibinfo{author}{\bibfnamefont{A.}~\bibnamefont{Kumar-Nayak}},
  \bibinfo{author}{\bibfnamefont{N.}~\bibnamefont{Morali}},
  \bibinfo{author}{\bibfnamefont{R.}~\bibnamefont{Batabyal}},
  \bibinfo{author}{\bibfnamefont{B.}~\bibnamefont{Yan}}, \bibnamefont{and}
  \bibinfo{author}{\bibfnamefont{H.}~\bibnamefont{Beidenkopf}},
  {``}\bibinfo{title}{Quasiparticle Interference Studies of Quantum
  Materials},{''} \bibinfo{journal}{Advanced Materials}
  \textbf{\bibinfo{volume}{30}}, \bibinfo{pages}{1707628}
  (\bibinfo{year}{2018}).

\bibitem[{\citenamefont{Hasan and Kane}(2010)}]{Hasan2010}
\bibinfo{author}{\bibfnamefont{M.~Z.} \bibnamefont{Hasan}} \bibnamefont{and}
  \bibinfo{author}{\bibfnamefont{C.~L.} \bibnamefont{Kane}},
  {``}\bibinfo{title}{Colloquium: Topological insulators},{''}
  \bibinfo{journal}{Rev. Mod. Phys.} \textbf{\bibinfo{volume}{82}},
  \bibinfo{pages}{3045} (\bibinfo{year}{2010}).

\bibitem[{\citenamefont{Akbari et~al.}(2010)\citenamefont{Akbari, Knolle,
  Eremin, and Moessner}}]{Akbari2010}
\bibinfo{author}{\bibfnamefont{A.}~\bibnamefont{Akbari}},
  \bibinfo{author}{\bibfnamefont{J.}~\bibnamefont{Knolle}},
  \bibinfo{author}{\bibfnamefont{I.}~\bibnamefont{Eremin}}, \bibnamefont{and}
  \bibinfo{author}{\bibfnamefont{R.}~\bibnamefont{Moessner}},
  {``}\bibinfo{title}{Quasiparticle interference in iron-based
  superconductors},{''} \bibinfo{journal}{Phys. Rev. B}
  \textbf{\bibinfo{volume}{82}}, \bibinfo{pages}{224506}
  (\bibinfo{year}{2010}).

\bibitem[{\citenamefont{Hirschfeld et~al.}(2015)\citenamefont{Hirschfeld,
  Altenfeld, Eremin, and Mazin}}]{Hirschfeld2015}
\bibinfo{author}{\bibfnamefont{P.~J.} \bibnamefont{Hirschfeld}},
  \bibinfo{author}{\bibfnamefont{D.}~\bibnamefont{Altenfeld}},
  \bibinfo{author}{\bibfnamefont{I.}~\bibnamefont{Eremin}}, \bibnamefont{and}
  \bibinfo{author}{\bibfnamefont{I.~I.} \bibnamefont{Mazin}},
  {``}\bibinfo{title}{Robust determination of the superconducting gap sign
  structure via quasiparticle interference},{''} \bibinfo{journal}{Phys. Rev.
  B} \textbf{\bibinfo{volume}{92}}, \bibinfo{pages}{184513}
  (\bibinfo{year}{2015}).

\bibitem[{\citenamefont{Lee et~al.}(2010)\citenamefont{Lee, Arovas, and
  Wu}}]{Lee2010}
\bibinfo{author}{\bibfnamefont{W.-C.} \bibnamefont{Lee}},
  \bibinfo{author}{\bibfnamefont{D.~P.} \bibnamefont{Arovas}},
  \bibnamefont{and} \bibinfo{author}{\bibfnamefont{C.}~\bibnamefont{Wu}},
  {``}\bibinfo{title}{Quasiparticle interference in the unconventional
  metamagnetic compound ${\text{Sr}}_{3}{\text{Ru}}_{2}{\text{O}}_{7}$},{''}
  \bibinfo{journal}{Phys. Rev. B} \textbf{\bibinfo{volume}{81}},
  \bibinfo{pages}{184403} (\bibinfo{year}{2010}).

\bibitem[{\citenamefont{Farrell et~al.}(2015)\citenamefont{Farrell, Beaudry,
  Franz, and Pereg-Barnea}}]{Farrell2015}
\bibinfo{author}{\bibfnamefont{A.}~\bibnamefont{Farrell}},
  \bibinfo{author}{\bibfnamefont{M.}~\bibnamefont{Beaudry}},
  \bibinfo{author}{\bibfnamefont{M.}~\bibnamefont{Franz}}, \bibnamefont{and}
  \bibinfo{author}{\bibfnamefont{T.}~\bibnamefont{Pereg-Barnea}},
  {``}\bibinfo{title}{Quasiparticle interference patterns in a topological
  superconductor},{''} \bibinfo{journal}{Phys. Rev. B}
  \textbf{\bibinfo{volume}{91}}, \bibinfo{pages}{134510}
  (\bibinfo{year}{2015}).

\bibitem[{\citenamefont{Böker et~al.}(2019)\citenamefont{Böker, Volkov,
  Hirschfeld, and Eremin}}]{Boker2019}
\bibinfo{author}{\bibfnamefont{J.}~\bibnamefont{Böker}},
  \bibinfo{author}{\bibfnamefont{P.~A.} \bibnamefont{Volkov}},
  \bibinfo{author}{\bibfnamefont{P.~J.} \bibnamefont{Hirschfeld}},
  \bibnamefont{and} \bibinfo{author}{\bibfnamefont{I.}~\bibnamefont{Eremin}},
  {``}\bibinfo{title}{Quasiparticle interference and symmetry of
  superconducting order parameter in strongly electron-doped iron-based
  superconductors},{''} \bibinfo{journal}{New Journal of Physics}
  \textbf{\bibinfo{volume}{21}}, \bibinfo{pages}{083021}
  (\bibinfo{year}{2019}).

\bibitem[{\citenamefont{He}(2017)}]{Chaocheng2017}
\bibinfo{author}{\bibfnamefont{C.}~\bibnamefont{He}},
  {``}\bibinfo{title}{Quasiparticle interference in two-dimensional topological
  crystalline superconductors},{''} \bibinfo{journal}{{EPL} (Europhysics
  Letters)} \textbf{\bibinfo{volume}{120}}, \bibinfo{pages}{27003}
  (\bibinfo{year}{2017}).

\bibitem[{\citenamefont{Iwaya et~al.}(2017)\citenamefont{Iwaya, Kohsaka, Okawa,
  Machida, Bahramy, Hanaguri, and Sasagawa}}]{Iwaya2017}
\bibinfo{author}{\bibfnamefont{K.}~\bibnamefont{Iwaya}},
  \bibinfo{author}{\bibfnamefont{Y.}~\bibnamefont{Kohsaka}},
  \bibinfo{author}{\bibfnamefont{K.}~\bibnamefont{Okawa}},
  \bibinfo{author}{\bibfnamefont{T.}~\bibnamefont{Machida}},
  \bibinfo{author}{\bibfnamefont{M.~S.} \bibnamefont{Bahramy}},
  \bibinfo{author}{\bibfnamefont{T.}~\bibnamefont{Hanaguri}}, \bibnamefont{and}
  \bibinfo{author}{\bibfnamefont{T.}~\bibnamefont{Sasagawa}},
  {``}\bibinfo{title}{Author Correction: Full-gap superconductivity in
  spin-polarised surface states of topological semimetal $\beta$-{PdBi}2},{''}
  \bibinfo{journal}{Nature Communications} \textbf{\bibinfo{volume}{8}}
  (\bibinfo{year}{2017}).

\bibitem[{\citenamefont{Gu et~al.}(2018)\citenamefont{Gu, Tang, Wan, Du, Yang,
  Yang, Wang, Lin, Zhu, and Wen}}]{Gu2018}
\bibinfo{author}{\bibfnamefont{Q.}~\bibnamefont{Gu}},
  \bibinfo{author}{\bibfnamefont{Q.}~\bibnamefont{Tang}},
  \bibinfo{author}{\bibfnamefont{S.}~\bibnamefont{Wan}},
  \bibinfo{author}{\bibfnamefont{Z.}~\bibnamefont{Du}},
  \bibinfo{author}{\bibfnamefont{X.}~\bibnamefont{Yang}},
  \bibinfo{author}{\bibfnamefont{H.}~\bibnamefont{Yang}},
  \bibinfo{author}{\bibfnamefont{Q.-H.} \bibnamefont{Wang}},
  \bibinfo{author}{\bibfnamefont{H.}~\bibnamefont{Lin}},
  \bibinfo{author}{\bibfnamefont{X.}~\bibnamefont{Zhu}}, \bibnamefont{and}
  \bibinfo{author}{\bibfnamefont{H.-H.} \bibnamefont{Wen}},
  {``}\bibinfo{title}{Sign-reversal superconducting gaps revealed by
  phase-referenced quasiparticle interference of impurity-induced bound states
  in
  $({\mathrm{Li}}_{1\ensuremath{-}x}{\mathrm{Fe}}_{x}){\mathrm{OHFe}}_{1\ensuremath{-}y}{\mathrm{Zn}}_{y}\mathrm{Se}$},{''}
  \bibinfo{journal}{Phys. Rev. B} \textbf{\bibinfo{volume}{98}},
  \bibinfo{pages}{134503} (\bibinfo{year}{2018}).

\bibitem[{\citenamefont{Wang et~al.}(2017)\citenamefont{Wang, Walkup, Derry,
  Scaffidi, Rak, Vig, Kogar, Zeljkovic, Husain, Santos et~al.}}]{Wang2017}
\bibinfo{author}{\bibfnamefont{Z.}~\bibnamefont{Wang}},
  \bibinfo{author}{\bibfnamefont{D.}~\bibnamefont{Walkup}},
  \bibinfo{author}{\bibfnamefont{P.}~\bibnamefont{Derry}},
  \bibinfo{author}{\bibfnamefont{T.}~\bibnamefont{Scaffidi}},
  \bibinfo{author}{\bibfnamefont{M.}~\bibnamefont{Rak}},
  \bibinfo{author}{\bibfnamefont{S.}~\bibnamefont{Vig}},
  \bibinfo{author}{\bibfnamefont{A.}~\bibnamefont{Kogar}},
  \bibinfo{author}{\bibfnamefont{I.}~\bibnamefont{Zeljkovic}},
  \bibinfo{author}{\bibfnamefont{A.}~\bibnamefont{Husain}},
  \bibinfo{author}{\bibfnamefont{L.~H.} \bibnamefont{Santos}},
  \bibnamefont{et~al.}, {``}\bibinfo{title}{Quasiparticle interference and
  strong electron{\textendash}mode coupling in the quasi-one-dimensional bands
  of Sr2RuO4},{''} \bibinfo{journal}{Nature Physics}
  \textbf{\bibinfo{volume}{13}}, \bibinfo{pages}{799} (\bibinfo{year}{2017}).

\bibitem[{\citenamefont{Zhou et~al.}(2009)\citenamefont{Zhou, Fang, Tsai, and
  Hu}}]{Zhou2009}
\bibinfo{author}{\bibfnamefont{X.}~\bibnamefont{Zhou}},
  \bibinfo{author}{\bibfnamefont{C.}~\bibnamefont{Fang}},
  \bibinfo{author}{\bibfnamefont{W.-F.} \bibnamefont{Tsai}}, \bibnamefont{and}
  \bibinfo{author}{\bibfnamefont{J.}~\bibnamefont{Hu}},
  {``}\bibinfo{title}{Theory of quasiparticle scattering in a two-dimensional
  system of helical Dirac fermions: Surface band structure of a
  three-dimensional topological insulator},{''} \bibinfo{journal}{Phys. Rev. B}
  \textbf{\bibinfo{volume}{80}}, \bibinfo{pages}{245317}
  (\bibinfo{year}{2009}).

\bibitem[{\citenamefont{Liu et~al.}(2010)\citenamefont{Liu, Qi, Zhang, Dai,
  Fang, and Zhang}}]{Liu2010}
\bibinfo{author}{\bibfnamefont{C.-X.} \bibnamefont{Liu}},
  \bibinfo{author}{\bibfnamefont{X.-L.} \bibnamefont{Qi}},
  \bibinfo{author}{\bibfnamefont{H.}~\bibnamefont{Zhang}},
  \bibinfo{author}{\bibfnamefont{X.}~\bibnamefont{Dai}},
  \bibinfo{author}{\bibfnamefont{Z.}~\bibnamefont{Fang}}, \bibnamefont{and}
  \bibinfo{author}{\bibfnamefont{S.-C.} \bibnamefont{Zhang}},
  {``}\bibinfo{title}{Model Hamiltonian for topological insulators},{''}
  \bibinfo{journal}{Phys. Rev. B} \textbf{\bibinfo{volume}{82}},
  \bibinfo{pages}{045122} (\bibinfo{year}{2010}).

\bibitem[{\citenamefont{Zhang et~al.}(2009)\citenamefont{Zhang, Liu, Qi, Dai,
  Fang, and Zhang}}]{Zhang2009}
\bibinfo{author}{\bibfnamefont{H.}~\bibnamefont{Zhang}},
  \bibinfo{author}{\bibfnamefont{C.-X.} \bibnamefont{Liu}},
  \bibinfo{author}{\bibfnamefont{X.-L.} \bibnamefont{Qi}},
  \bibinfo{author}{\bibfnamefont{X.}~\bibnamefont{Dai}},
  \bibinfo{author}{\bibfnamefont{Z.}~\bibnamefont{Fang}}, \bibnamefont{and}
  \bibinfo{author}{\bibfnamefont{S.-C.} \bibnamefont{Zhang}},
  {``}\bibinfo{title}{Topological insulators in Bi2Se3, Bi2Te3 and Sb2Te3 with
  a single Dirac cone on the surface},{''} \bibinfo{journal}{Nature Physics}
  \textbf{\bibinfo{volume}{5}}, \bibinfo{pages}{438} (\bibinfo{year}{2009}).

\bibitem[{\citenamefont{Hao and Ting}(2017)}]{Hao2017}
\bibinfo{author}{\bibfnamefont{L.}~\bibnamefont{Hao}} \bibnamefont{and}
  \bibinfo{author}{\bibfnamefont{C.~S.} \bibnamefont{Ting}},
  {``}\bibinfo{title}{Nematic superconductivity in
  ${\mathrm{Cu}}_{x}{\mathrm{Bi}}_{2}{\mathrm{Se}}_{3}$: Surface Andreev bound
  states},{''} \bibinfo{journal}{Phys. Rev. B} \textbf{\bibinfo{volume}{96}},
  \bibinfo{pages}{144512} (\bibinfo{year}{2017}).

\bibitem[{\citenamefont{Beidenkopf et~al.}(2011)\citenamefont{Beidenkopf,
  Roushan, Seo, Gorman, Drozdov, Hor, Cava, and Yazdani}}]{Beidenkopf2011}
\bibinfo{author}{\bibfnamefont{H.}~\bibnamefont{Beidenkopf}},
  \bibinfo{author}{\bibfnamefont{P.}~\bibnamefont{Roushan}},
  \bibinfo{author}{\bibfnamefont{J.}~\bibnamefont{Seo}},
  \bibinfo{author}{\bibfnamefont{L.}~\bibnamefont{Gorman}},
  \bibinfo{author}{\bibfnamefont{I.}~\bibnamefont{Drozdov}},
  \bibinfo{author}{\bibfnamefont{Y.~S.} \bibnamefont{Hor}},
  \bibinfo{author}{\bibfnamefont{R.~J.} \bibnamefont{Cava}}, \bibnamefont{and}
  \bibinfo{author}{\bibfnamefont{A.}~\bibnamefont{Yazdani}},
  {``}\bibinfo{title}{Spatial fluctuations of helical Dirac fermions on the
  surface of topological insulators},{''} \bibinfo{journal}{Nature Physics}
  \textbf{\bibinfo{volume}{7}}, \bibinfo{pages}{939} (\bibinfo{year}{2011}).

\bibitem[{\citenamefont{Kohsaka et~al.}(2017)\citenamefont{Kohsaka, Machida,
  Iwaya, Kanou, Hanaguri, and Sasagawa}}]{Kohsaka2017}
\bibinfo{author}{\bibfnamefont{Y.}~\bibnamefont{Kohsaka}},
  \bibinfo{author}{\bibfnamefont{T.}~\bibnamefont{Machida}},
  \bibinfo{author}{\bibfnamefont{K.}~\bibnamefont{Iwaya}},
  \bibinfo{author}{\bibfnamefont{M.}~\bibnamefont{Kanou}},
  \bibinfo{author}{\bibfnamefont{T.}~\bibnamefont{Hanaguri}}, \bibnamefont{and}
  \bibinfo{author}{\bibfnamefont{T.}~\bibnamefont{Sasagawa}},
  {``}\bibinfo{title}{Spin-orbit scattering visualized in quasiparticle
  interference},{''} \bibinfo{journal}{Phys. Rev. B}
  \textbf{\bibinfo{volume}{95}}, \bibinfo{pages}{115307}
  (\bibinfo{year}{2017}).

\bibitem[{\citenamefont{Neupane et~al.}(2016)\citenamefont{Neupane, Ishida,
  Sankar, Zhu, Sanchez, Belopolski, Xu, Alidoust, Hosen, Shin
  et~al.}}]{Neupane2016}
\bibinfo{author}{\bibfnamefont{M.}~\bibnamefont{Neupane}},
  \bibinfo{author}{\bibfnamefont{Y.}~\bibnamefont{Ishida}},
  \bibinfo{author}{\bibfnamefont{R.}~\bibnamefont{Sankar}},
  \bibinfo{author}{\bibfnamefont{J.-X.} \bibnamefont{Zhu}},
  \bibinfo{author}{\bibfnamefont{D.~S.} \bibnamefont{Sanchez}},
  \bibinfo{author}{\bibfnamefont{I.}~\bibnamefont{Belopolski}},
  \bibinfo{author}{\bibfnamefont{S.-Y.} \bibnamefont{Xu}},
  \bibinfo{author}{\bibfnamefont{N.}~\bibnamefont{Alidoust}},
  \bibinfo{author}{\bibfnamefont{M.~M.} \bibnamefont{Hosen}},
  \bibinfo{author}{\bibfnamefont{S.}~\bibnamefont{Shin}}, \bibnamefont{et~al.},
  {``}\bibinfo{title}{Electronic structure and relaxation dynamics in a
  superconducting topological material},{''} \bibinfo{journal}{Scientific
  Reports} \textbf{\bibinfo{volume}{6}}, \bibinfo{pages}{22557}
  (\bibinfo{year}{2016}).

\bibitem[{\citenamefont{Lee et~al.}(2009)\citenamefont{Lee, Wu, Arovas, and
  Zhang}}]{Lee2009}
\bibinfo{author}{\bibfnamefont{W.-C.} \bibnamefont{Lee}},
  \bibinfo{author}{\bibfnamefont{C.}~\bibnamefont{Wu}},
  \bibinfo{author}{\bibfnamefont{D.~P.} \bibnamefont{Arovas}},
  \bibnamefont{and} \bibinfo{author}{\bibfnamefont{S.-C.} \bibnamefont{Zhang}},
  {``}\bibinfo{title}{Quasiparticle interference on the surface of the
  topological insulator ${\text{Bi}}_{2}{\text{Te}}_{3}$},{''}
  \bibinfo{journal}{Phys. Rev. B} \textbf{\bibinfo{volume}{80}},
  \bibinfo{pages}{245439} (\bibinfo{year}{2009}).

\bibitem[{\citenamefont{Bao et~al.}(2018)\citenamefont{Bao, Tang, Lu, and
  Wang}}]{Bao2018}
\bibinfo{author}{\bibfnamefont{W.-C.} \bibnamefont{Bao}},
  \bibinfo{author}{\bibfnamefont{Q.-K.} \bibnamefont{Tang}},
  \bibinfo{author}{\bibfnamefont{D.-C.} \bibnamefont{Lu}}, \bibnamefont{and}
  \bibinfo{author}{\bibfnamefont{Q.-H.} \bibnamefont{Wang}},
  {``}\bibinfo{title}{Visualizing the $d$ vector in a nematic triplet
  superconductor},{''} \bibinfo{journal}{Phys. Rev. B}
  \textbf{\bibinfo{volume}{98}}, \bibinfo{pages}{054502}
  (\bibinfo{year}{2018}).

\bibitem[{\citenamefont{Levy et~al.}(2013)\citenamefont{Levy, Zhang, Ha,
  Sharifi, Talin, Kuk, and Stroscio}}]{Levy2013}
\bibinfo{author}{\bibfnamefont{N.}~\bibnamefont{Levy}},
  \bibinfo{author}{\bibfnamefont{T.}~\bibnamefont{Zhang}},
  \bibinfo{author}{\bibfnamefont{J.}~\bibnamefont{Ha}},
  \bibinfo{author}{\bibfnamefont{F.}~\bibnamefont{Sharifi}},
  \bibinfo{author}{\bibfnamefont{A.~A.} \bibnamefont{Talin}},
  \bibinfo{author}{\bibfnamefont{Y.}~\bibnamefont{Kuk}}, \bibnamefont{and}
  \bibinfo{author}{\bibfnamefont{J.~A.} \bibnamefont{Stroscio}},
  {``}\bibinfo{title}{Experimental Evidence for $s$-Wave Pairing Symmetry in
  Superconducting ${\mathrm{Cu}}_{x}{\mathrm{Bi}}_{2}{\mathrm{Se}}_{3}$ Single
  Crystals Using a Scanning Tunneling Microscope},{''} \bibinfo{journal}{Phys.
  Rev. Lett.} \textbf{\bibinfo{volume}{110}}, \bibinfo{pages}{117001}
  (\bibinfo{year}{2013}).

\bibitem[{\citenamefont{Peng et~al.}(2013)\citenamefont{Peng, De, Lv, Wei, and
  Chu}}]{Peng2013}
\bibinfo{author}{\bibfnamefont{H.}~\bibnamefont{Peng}},
  \bibinfo{author}{\bibfnamefont{D.}~\bibnamefont{De}},
  \bibinfo{author}{\bibfnamefont{B.}~\bibnamefont{Lv}},
  \bibinfo{author}{\bibfnamefont{F.}~\bibnamefont{Wei}}, \bibnamefont{and}
  \bibinfo{author}{\bibfnamefont{C.-W.} \bibnamefont{Chu}},
  {``}\bibinfo{title}{Absence of zero-energy surface bound states in
  Cu${}_{x}$Bi${}_{2}$Se${}_{3}$ studied via Andreev reflection
  spectroscopy},{''} \bibinfo{journal}{Phys. Rev. B}
  \textbf{\bibinfo{volume}{88}}, \bibinfo{pages}{024515}
  (\bibinfo{year}{2013}).

\bibitem[{\citenamefont{Hsieh and Fu}(2012)}]{Hsieh2012}
\bibinfo{author}{\bibfnamefont{T.~H.} \bibnamefont{Hsieh}} \bibnamefont{and}
  \bibinfo{author}{\bibfnamefont{L.}~\bibnamefont{Fu}},
  {``}\bibinfo{title}{Majorana Fermions and Exotic Surface Andreev Bound States
  in Topological Superconductors: Application to
  ${\mathrm{Cu}}_{x}{\mathrm{Bi}}_{2}{\mathrm{Se}}_{3}$},{''}
  \bibinfo{journal}{Phys. Rev. Lett.} \textbf{\bibinfo{volume}{108}},
  \bibinfo{pages}{107005} (\bibinfo{year}{2012}).

\bibitem[{\citenamefont{Hao and Lee}(2015)}]{Hao2015}
\bibinfo{author}{\bibfnamefont{L.}~\bibnamefont{Hao}} \bibnamefont{and}
  \bibinfo{author}{\bibfnamefont{T.-K.} \bibnamefont{Lee}},
  {``}\bibinfo{title}{Effective low-energy theory for superconducting
  topological insulators},{''} \bibinfo{journal}{Journal of Physics: Condensed
  Matter} \textbf{\bibinfo{volume}{27}}, \bibinfo{pages}{105701}
  (\bibinfo{year}{2015}).

\end{thebibliography}
\end{document}